\documentclass[a4paper,11pt]{article}
\usepackage{jcappub} % for details on the use of the package, please see the JINST-author-manual
\usepackage{lineno}

\newcommand{\lcdm}{$\Lambda$CDM}
\newcommand{\Neff}{$N_\textrm{eff}$}
\newcommand{\Planck}{{\em Planck}}
\newcommand{\SO}{{\em Simons Observatory}}
\newcommand{\Oedf}{\Omega_\text{EDF}}

\title{A flexible parameterization to test early physics solutions to the Hubble tension with future CMB data}

\author[1]{Raphaël Kou\note{Corresponding author.}}
\author{and Antony Lewis}
\affiliation{Department of Physics \& Astronomy, University of Sussex, Brighton BN1 9QH, UK}

\emailAdd{r.kou@sussex.ac.uk}

\abstract{

One approach to reconciling local measurements of a high expansion rate with observations of acoustic oscillations in the CMB and galaxy clustering (the ``Hubble tension'') is to introduce additional contributions to the \lcdm\ model that are relevant before recombination. While numerous possibilities exist, none are currently well-motivated or preferred by data. However, future CMB experiments, which will measure acoustic peaks to much smaller scales and resolve polarization signals with higher signal-to-noise ratio over large sky areas, should detect almost any such modification at high significance.

We propose a method to capture most relevant possible deviations from \lcdm\ due to additional non-interacting components, while remaining sufficiently constraining to enable detection across various scenarios. The phenomenological model uses a fluid model with four parameters governing additional density contributions that peak at different redshifts, and two sound speed parameters. We forecast possible constraints with \SO, explore parameter degeneracies that arise in \lcdm, and demonstrate that this method could detect a range of specific models. Which of the new parameters gets excited can give hints about the nature of any new physics, while the generality of the model allows for testing with future data in a way that should not be plagued by a posteriori choices and would reduce publication bias.

When testing our model with \Planck\ data, we find good consistency with the \lcdm\ model, but the data also allows for a large Hubble parameter, especially if the sound speed of an additional component is not too different from that of radiation. The analysis with \Planck\ data reveals significant volume effects, requiring careful interpretation of results. We demonstrate that \SO\ data will mitigate these volume effects, so that any indicated solution to the Hubble tension using our model cannot be mimicked by volume effects alone, given the significance of the tension.
}

\begin{document}
\maketitle
\flushbottom

\section{Introduction}
\label{sec:intro}

A major puzzle in modern cosmology is the persistent discrepancy between local measurements of the Hubble constant \(H_0\) and the value inferred from cosmological observations assuming a standard $\Lambda$--Cold Dark Matter (\lcdm) model. This is commonly known as the Hubble tension (for a comprehensive review, including potential solutions, see ref.~\cite{Divalentino21,Abdalla22}). 
%Despite improved observational accuracy, data from the Cosmic Microwave Background (CMB) and some local distance-ladder methods continue to yield conflicting values for \(H_0\), suggesting the need for new physics beyond the standard \lcdm\ model. 
The distance-ladder method involves successive calibrations of the luminosity of bright objects, primarily Cepheids and Type Ia supernovae (SNe Ia), with the latest analysis from the SH0ES team~\cite{Breuval:2024lsv} giving the constraint $H_0 = 73.17\, \pm\, 0.86\,\rm{km}\,\rm{s}^{-1}\,\rm{Mpc}^{-1}$. Cosmological constraints are mainly derived from observations of the cosmic microwave background (CMB), where the $H_0$ value is indirectly inferred by fitting an assumed theoretical model. In the \lcdm\ model, the PR3 release from the \Planck\ collaboration~\cite{Planck18} gives $H_0 = 67.27\, \pm\, 0.60\,\rm{km}\,\rm{s}^{-1}\,\rm{Mpc}^{-1}$, in good agreement with more recent measurements and results from alternative inverse-distance ladders using baryonic acoustic oscillation (BAO) data combined with constraints on the baryon density~\cite{Aubourg15,Macaulay19,DESI24,Camilleri24}.
Comparing these constraints, the Hubble tension is at the $>5\sigma$ level. However, some analyses of different local distance ladders have recently given local measurements more consistent with the \Planck\ value, with ongoing debate about the details of the analysis~\cite{Freedman:2024eph,Riess:2024vfa}. 
As an indication of new physics, it would be much clearer if the new physics could be detected from CMB data alone. The main goal of this paper is to suggest a way to test new early physics from future CMB data in a flexible and relatively general way, avoiding issues of a posteriori choices that can result from deciding on which models to test only after the new data is available.

One potential solution to the tension is to allow deviations from \lcdm\ in the early universe. The CMB anisotropies constrain the observed angular acoustic scale $\theta_* = r_*/D(z_*)$ to high ($\sim 0.05\%$, see~\cite{Planck18}) accuracy in a fairly model-independent way, where $r_*$ is the comoving sound horizon at recombination and $D(z_*)$ is the comoving angular diameter distance to recombination ($z_*$ denotes the redshift of recombination). This however leaves significant freedom to vary $r_*$ and $D(z_*)$ as long as their observed ratio remains nearly constant. One possible solution to the Hubble tension would be new physics that decreases $r_*$ and $D(z_*)$, with a decreased $D(z_*)$ implying a larger $H_0$ if the late-time evolution remains close to \lcdm.

Among many models that have been suggested, the Early Dark Energy (EDE) scenario appeared as an interesting candidate to solve the Hubble tension (see~\cite{Poulin23} for a full review). EDE models usually assume the existence of a scalar field that will contribute significantly to the background expansion at some point in the Universe's history around redshift $z~\sim~10^3~-~10^4$. More precisely, those models generally assume the field to be frozen at very high redshift, so that its energy density is constant, and its fractional contribution to the expansion rate then increases over time up to the point when the field starts to evolve and its density dilutes faster than matter. These models gained much attention in the context of the Hubble tension, particularly after ref.~\cite{Poulin19} suggested they might be able to reconcile early and late time observations. Certain analyses of the \textit{South Pole Telescope} (SPT) and the \textit{Atacama Cosmology Telescope} (ACT) data have found moderate preference for EDE over \lcdm\ without including any $H_0$ prior, primarily driven by the latter data set. The ability of these models to increase the inferred value of $H_0$ comes from the fact that adding an extra component to the background density increases the early expansion rate, decreasing the time to recombination, and therefore decreasing the sound horizon $r_*$. However, none are very compelling from a theoretical point of view, and few would be favoured from CMB data alone in the absence of the Hubble tension.

A number of other potential solutions to the Hubble tension involves additional ultra-relativistic species that interact weakly with ordinary matter and can constitute a \textit{dark radiation}~\cite{Hou13,Akita20,Froustey20,Bennett21,Gariazzo23}. The dark radiation energy density is usually modelled by adding extra relativistic degrees of freedom so that the radiation density becomes,
\begin{align}\label{eq:Neff}
    \rho_R=\rho_\gamma\left(1+\frac{7}{8}\left(\frac{4}{11}\right)^{4/3}N_\textrm{eff}\right),
\end{align}
where $\rho_\gamma$ is the photon energy density and $N_\textrm{eff}$ is the effective neutrino number used to quantify the amount of additional neutrinos or any other non-interacting relativistic species. Adding new species would lead to $N_\textrm{eff}>3.044$, the value predicted by the Standard Model of particle physics. Other dark radiation models also consider self-interacting dark radiation~\cite{Jeong13,Cyr16,Lesgourgues16,Khalife24} which differs from the free-streaming (non-interacting) case described above, as at the perturbation level the self-interaction can affect the clustering. As in the case of EDE, increasing the background energy density of the Universe with a new component would decrease the sound horizon and lead to the inference of a higher $H_0$ value. As a result, several studies have shown that dark radiation models can ease the Hubble tension by partially reconciling the discrepancy between local and early Universe measurements of $H_0$~\cite{Ghosh21,Aloni22,Brinckmann23,Bagherian24,Allali24}.

However, recent data analyses do not especially favour EDE or dark radiation as solutions of the Hubble tension~\cite{Efstathiou24,Khalife24,Hill20,McDonough24} and there is no strong evidence for either of those two classes of models, though unidentified systematic errors in the current data cannot be ruled out. A more general approach such as the Generalized Dark Matter (GDM)~\cite{Hu98} can also be considered. With that approach, one can model a very wide variety of extra non-interacting fluids by choosing a set of three functions. At the background level, the equation of state parameter $w=p/\rho$ relates the pressure $p$ to the density $\rho$ of the fluid. Modelling the perturbations then requires specification of the pressure perturbation (isotropic stress) $\delta p$ and the anisotropic stress $\Pi$. Using this approach, it is therefore possible to probe the properties of an extra fluid without assuming a specific theoretical model by directly parameterizing the equation of state, isotropic and anisotropic stress of the fluid. 

Ongoing and future CMB experiments, such as SPT~\cite{Ruhl04,Carlstrom11,Benson14,Sobrin22,Pan23,Balkenhol23,Prabhu24,Ge24}, ACT (whose data analysis continues, although the telescope has been decommissioned)~\cite{Fowler07,Thornton16,Henderson16,Aiola20,Choi20,Qu24,Madhavacheril24}, \SO~\cite{SimonsObs} or \textit{CMB-S4}~\cite{Abazajian16} will have increasing precision at small scales and in polarization, which will allow them to detect potential deviations from \lcdm\ without relying on late-time measurements of \(H_0\). In this context, we aim to develop a general and flexible parameterization of the early Universe relying on the GDM framework, specifically tailored to be tested with \SO\ data. This parameterization is designed to approximately reproduce existing theoretical models, such as EDE or dark radiation, but is not limited to them, offering the potential to uncover new physics from future CMB data. More precisely, our model introduces a new fluid in the early Universe that we call \textit{Early Dark Fluid} (EDF) and we parameterize its density and isotropic stress through the fluid's sound speed (more details in section~\ref{sec:sound_speed}). We neglect the anisotropic stress since we found that in simple cases varying the anisotropic stress did not help to better reproduce existing theoretical models, but rather introduced more degeneracies between parameters that \SO\ would not be able to disentangle. This approach was (at least partially) followed by a number of studies~\cite{Hojjati13,Moss21,Meiers23} that also parameterized the density and sound speed of an additional fluid. There is however a trade-off between having more parameters, allowing high-fidelity reproduction of specific models, and the resulting high-dimensional space where many directions are poorly constrained, making it difficult to separate random fluctuations from actual deviations from the \lcdm\ model.
Our approach builds upon previous studies by providing a way to estimate a reduced set of physically allowed combinations of density parameters which will be best constrained by \SO\ and that we call \textit{modes}, by analogy to the principal component analysis (PCA). We also carefully study how our parameterization is able to reproduce some existing theoretical models, and test whether \SO\ will be able to discriminate reliably between those models and \lcdm.

In section~\ref{sec:fluid_modelling}, we introduce the underlying model and estimate the best constrained density modes. We show in section~\ref{sec:test_models} how, using those modes, we can capture most of the effect of differences between models at the level of the CMB temperature and polarization power spectra. This approach is a way of testing the \lcdm\ model and can also provide hints about what kind of new physics is required in the early Universe. Finally, we constrain our model using current \Planck\ data in section~\ref{sec:Planck_constraints} to evaluate the state of the art.

\section{Early Dark Fluid (EDF) modelling}
\label{sec:fluid_modelling}

\subsection{Underlying density modelling}

The addition of a new Early Dark Fluid (EDF) modifies the background expansion of the Universe following the Friedmann equation as:
\begin{align}\label{eq:Friedmann}
    H^2(a) = H_0^2\left[\Omega_{\Lambda\textrm{CDM}}(a) + \Oedf(a)\right],
\end{align}
where $H(a)$ is the Hubble rate at a scale factor $a$, $H_0$ is the Hubble rate today ($a=1$), $\Omega_{\Lambda\textrm{CDM}}$ is the total \lcdm\ density (matter, radiation and cosmological constant) and $\Oedf(a)$ is the density parameter of the newly introduced fluid (the density at scale factor $a$ relative to the critical density today). We adopt the parameterization from ref.~\cite{Moss21} for $\Oedf$, such that,
\begin{align}
    \Oedf(a) &= \sum_{i=1}^{N}\Omega_i(a) \\
    \Omega_i(a) &= \Omega_i\Omega_{\Lambda\textrm{CDM}}(a_i)\left(\frac{2a_i^\beta}{a^\beta+a_i^\beta}\right)^{6/\beta},
\end{align}
where $\Omega_i$ is a free parameter and $\beta$ is a constant fixed to $\beta=6$. The EDF is therefore modelled as a set of $N$ fluids, each with a density $\Omega_i(a)$ that peaks relative to the background density at $a=a_i$. The use of this parameterization is particularly interesting because each of the $N$ components has an equation of state such that, 
\begin{align}
    -1\leq w_i(a) = \frac{2}{1+(a_i/a)^\beta}-1 \leq 1,
\end{align}
which ensures that the total equation of state of the EDF is also bounded between $-1$ and $1$ as long as each amplitude $\Omega_i$ is positive. As stated in ref.~\cite{Moss21}, those $N$ fluids can be seen as a well-defined set of basis functions that we use to parameterize the overall EDF density. The particular choice for the shape of the density evolution with scale factor and $\beta=6$ was made in order to ensure rapid decay to low redshift (to have sufficiently small early Integrated Sachs Wolfe (ISW) effect to be compatible with the current data), and reproduce the rapid decay of an axion fluid (see eq.~16 from ref.~\cite{Poulin18} with $w_n=1$).

We adopt a set of $N=50$ underlying $a_i$ values, logarithmically spaced between $a_1=10^{-6}$ and $a_N=0.1$. The lower bound was chosen due to the limited sensitivity of CMB data at smaller scale factors, and we avoid larger scale factors as the CMB is less sensitive to late-time physics. We do not consider late-time solutions to the $H_0$ tension in this paper, assuming standard \lcdm\ evolution at low redshift.

Those equations describe the background density of the EDF and its effect on the expansion of the Universe through the Friedmann equation (see eq.~\ref{eq:Friedmann}). At the level of the perturbations, the equations for the dark fluid perturbations in the synchronous gauge follow~\cite{Ma95,Weller03},
\begin{align}\label{eq:perturbations1}
    \delta'+3\mathcal{H}(\overline{c}_s^2-w)(\delta+3\mathcal{H}(1+w)v/k)+(1+w)kv+3\mathcal{H}w'v/k &=-(1+w)\frac{h'}{2} \\
    \label{eq:perturbations2}
    v' +(1-3\overline{c}_s^2)\mathcal{H}v &=\frac{k\overline{c}_s^2\delta}{1+w},
\end{align}
where $\delta$ is the density contrast, $\mathcal{H}=aH$ is the conformal Hubble parameter, $\overline{c}_s^2$ is the sound speed of the fluid in its rest frame, $v$ is the velocity, $k$ is the mode and $h$ is the trace part of the metric perturbation. Derivatives, denoted by prime symbols ($'$), are taken with respect to the conformal time and we impose null initial conditions.

\subsection{Best constrained modes}

In order to reduce the number of free parameters, we want to use only a combination of density parameters that is well constrained by the observations (here the CMB temperature and polarization anisotropies). The most common method to reduce the number of parameters is the Principal Component Analysis (PCA). However, some of the eigenmodes derived from PCA take negative values, preventing their amplitudes from being used as free parameters in a fitting process, as this could result in unphysical negative density values. Using modes that go negative would require using enough eigenvectors to ensure the summed model is physical, as well as very contrived priors, making the analysis of the posteriors tricky. By analogy to the PCA, we instead aim to find the ``eigenvectors" or ``modes" which are best constrained by the observations under the constraint that the modes only take positive values.

\sloppy Let us consider one observable $C_\ell$, which in our case is the concatenation of the CMB temperature and E polarization power spectra and cross-correlation, such that $C_\ell = (C_\ell^{TT},C_\ell^{EE},C_\ell^{TE})$. This observable can be modelled using a number of parameters, namely the set of $N$ amplitudes $\Omega_i$ and the \lcdm\ cosmological parameters that we denote $\theta=(\theta_*,\omega_b,\omega_c,\tau,A_se^{-2\tau},n_s)$ where $\theta_*$ is the CMB acoustic scale, $\omega_b\equiv \Omega_b h^2$ is the relative baryon density, $\omega_c\equiv \Omega_c h^2$ is the cold dark matter density, $\tau$ is the reionization optical depth, and $A_s$ and $n_s$ are respectively the amplitude and the tilt of the primordial power spectrum. We can then write,
\begin{align}\label{eq:cl_theta_omega}
    C_\ell = C_\ell(\theta,\Omega_1,\dots ,\Omega_N)\,.
\end{align}
We now describe the methodology we used to estimate the modes that we denote $(A_1, A_2, \dots, A_N)$, where all the modes are vectors of length $N$. For each mode $A_k$, we further write $A_k=\begin{pmatrix}
    A_{k1} & A_{k2} & \dots & A_{kN}
\end{pmatrix}^T$. If the modes are linearly independent, they then form a basis such that for any set of amplitudes $(\Omega_1,\dots,\Omega_N)$, there exist scalars $(x_1,\dots,x_N)$ such that,
\begin{align}
    \begin{pmatrix}
        \Omega_1 \\
        \Omega_2 \\
        \dots \\
        \Omega_N
    \end{pmatrix}
    =
    \sum_{k=1}^N x_kA_k
    = 
    \sum_{k=1}^N x_k
    \begin{pmatrix}
        A_{k1} \\
        A_{k2} \\
        \dots \\
        A_{kN}
    \end{pmatrix}.
\end{align}
Because our aim is to reduce the dimensionality of the parameter space, instead of estimating $N$ modes, we will only estimate $P$ modes ($P<N$) and use the approximation,
\begin{align}\label{eq:approx_PCA}
    \begin{pmatrix}
        \Omega_1 \\
        \Omega_2 \\
        \dots \\
        \Omega_N
    \end{pmatrix}
    \approx
    \sum_{k=1}^P x_kA_k,
\end{align}
which is similar to how PCA reduces dimensionality. We estimate the modes in sequential order: we first estimate $A_1$, then $A_2$, up to $A_P$. This is not optimal but, as shown below, estimating a mode will imply performing a minimization on $N$ parameters so that estimating the $P$ modes simultaneously would require to perform a minimization with $NP$ parameters, which is impractical in our case given that we have $N=50$. Assuming we have already estimated $k-1$ modes, here is how we estimate the $k^\textrm{th}$ mode. Using equations~\ref{eq:cl_theta_omega} and~\ref{eq:approx_PCA}, we can write,
\begin{align}
    \frac{dC_\ell}{dx_k} &= \sum_{i=1}^N\frac{dC_\ell}{d\Omega_i}\frac{d\Omega_i}{dx_k} \\
    &= \sum_{i=1}^N\frac{dC_\ell}{d\Omega_i}A_{ki}\,.
\end{align}
We can then estimate the 1-$\sigma$ uncertainty of the $x_k$ amplitude using the Fisher formalism. We recall that for two parameters $\alpha$ and $\beta$, which in our case are either one of the six \lcdm\ cosmological parameters or one of the $k$ amplitudes $(x_1,\dots,x_k)$ considered up to this point, the Fisher matrix reads,
\begin{align}\label{eq:Fisher}
    F_{\alpha,\beta} = \left(\frac{\partial C_\ell}{\partial \alpha}\right)^T\mathcal{C}^{-1}\frac{\partial C_\ell}{\partial \beta},
\end{align}
where $\mathcal{C}$ is the covariance matrix of the observations. The 1-$\sigma$ uncertainty in $x_k$ can be estimated after marginalizing over the \lcdm\ parameters and the previous $k-1$ amplitudes $(x_1,\dots,x_{k-1})$, as follows,
\begin{align}\label{eq:sigma_x}
    \sigma(x_k) = \left(F^{-1}\right)_{x_k,x_k}\,.
\end{align}
In order to estimate the $k^\textrm{th}$ positive mode that is best constrained by the observations, we simultaneously vary the $N$ coefficients of $A_k$ in order to minimize eq.~\ref{eq:sigma_x} under the constraint that all $A_{ki}$ take positive values and that $A_k$ is normalized (i.e.~$\lVert~A_k~\rVert_2^2~=~\sum_{i=1}^N~A_{ki}^2~=~1$). Performing this minimization therefore allows us to estimate the $k^\textrm{th}$ mode $A_k$, before starting the estimation of $A_{k+1}$. As noted previously, the optimal approach would be to estimate all modes simultaneously rather than sequentially, but this is in practice very difficult if $N$ is a high number.

In practice, we use CAMB~\cite{camb,Howlett:2012mh} to model the CMB power spectra, focusing on the unlensed predictions to ensure that the estimated modes reflect early Universe physics, unaffected by late-time lensing. The fiducial cosmology that we use corresponds to the (TT,TE,EE + lowE + lensing + BAO) column from ref.~\cite{Planck18} and we take all the amplitudes to be $\Omega_i=0.01$. Not using a \lcdm\ fiducial cosmology allows for a more stable evaluation of the derivatives $\frac{dC_\ell}{d\Omega_i}$, though the modes we estimate can then depend on the fiducial model to some extent. The covariance matrix $\mathcal{C}$ used in eq.~\ref{eq:Fisher} is a Gaussian covariance matrix~\cite{Tristram05} following,

\begin{align}
    \mathcal{C}(C_\ell^{AB},C_{\ell'}^{CD}) = \frac{\delta_{\ell\ell'}}{(2\ell+1)f_\textrm{sky}}\left[\left(C_\ell^{AC}+N_\ell^{AC}\right)\left(C_\ell^{BD}+N_\ell^{BD}\right)+\left(C_\ell^{BC}+N_\ell^{BC}\right)\left(C_\ell^{AD}+N_\ell^{AD}\right)\right]
\end{align}
where $A,B,C,D$ denote either temperature or E polarization, $f_\textrm{sky}$ is the observed sky fraction for which we use $40\%$ and the different $N_\ell$ are the foreground-cleaned, beam-corrected noise curves provided by \SO~\cite{SimonsObs} at baseline level. We consider multipoles in the range $\ell \in [40,5000]$. Because \Planck\ observed about $70\%$ of the sky, it will be possible to use additional observations from \Planck\ in areas that \SO\ will not cover. 
To construct a model tailored to \SO’s constraining power, we do not take into account the remaining $30\%$, but including it could tighten the constraints. \Planck\ was also able to measure power spectra at multipoles below $\ell<40$. We do not include these low multipoles, but we do add a prior on the reionization optical depth $\tau$ comparable to that obtained using CMB polarization observations at low multipoles. In practice, we include a Gaussian prior with standard deviation $\sigma=0.0073$ by adding $1/\sigma^2$ to the component corresponding to $\tau$ in the Fisher matrix computed with eq.~\ref{eq:Fisher}.

Following this approach, we estimated sequentially ten modes and eventually kept only the first four in order to reduce the degeneracies that can appear when using many modes. On top of that, higher modes quickly started to look wiggly (in scale factor space), which may be due to the fact that they were estimated sequentially, and that most of the information is contained in the first modes. To quantify the information contained in a subset of parameters, we evaluate the sum of the eigenvalues of the Fisher matrix corresponding to the selected parameters. Specifically, we denote by $F_k$ the Fisher matrix of the first $k$ modes (marginalized over the \lcdm\ parameters) and compute:
\begin{align}
    S_k = \sum_{j=1}^k \lambda_{kj},
\end{align}
where $\lambda_{kj}$ are the eigenvalues of $F_k$. Using this methodology, we find $S_4/S_{10}=0.68$, indicating that the first four modes contain 
$68\%$ of the information available in the ten modes. This ratio even reaches $0.78$ after marginalizing over the sound speed parameters (see section~\ref{sec:sound_speed}). This approach generalizes the cumulative inverse variance method of ref.~\cite{Samsing12}, which is applicable to diagonal Fisher matrices, such as those obtained through PCA. While the higher modes may still contain useful information, including a larger number of modes would likely increase degeneracies and hinder the convergence of Monte Carlo Markov Chains (MCMC) within a reasonable computational time.

\begin{figure}[htbp]
\centering
\includegraphics[width=\textwidth]{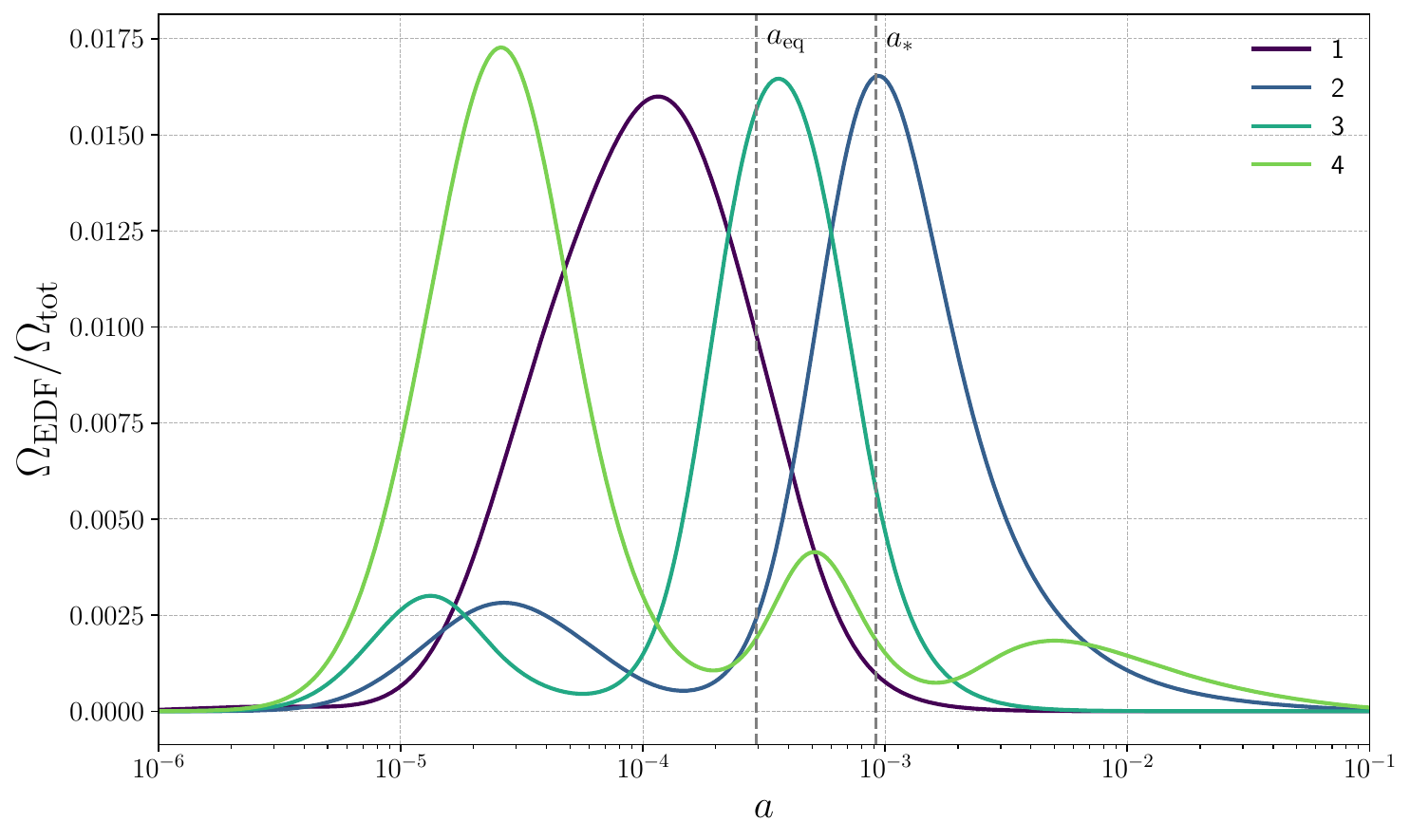}
\caption{Ratio of the EDF density to the total (i.e. \lcdm\ and EDF) density, where the EDF density corresponds to the positive modes which are best constrained by the CMB. The amplitude of the modes has arbitrarily been fixed to $0.01$ for the plot. The number $i$ indicated in legend means that the corresponding mode is the $i^\textrm{th}$ best constrained by the observations. The grey dashed lines show the matter-radiation equality and recombination scale factors for reference.\label{fig:modes}}
\end{figure}

Figure~\ref{fig:modes} shows those four density modes as a function of scale factor. Even if they look quite simple, the approach we followed allocates the location and width of the modes in a less arbitrary way than randomly selecting a number of scale factor bins. Those modes can be compared to the ones obtained by ref.~\cite{Samsing12} who performed the standard PCA to estimate the modes best constrained by CMB observations (without the positivity constraint). The two sets of modes share the qualitative trait of being quite broad and not oscillating too much, contrary to the higher, unconstrained, principal components of~\cite{Samsing12}. It appears, however, that our modes span a wider range of scale factors, roughly from $a=10^{-5}$ to $a=10^{-2}$ whereas that of~\cite{Samsing12} are more localized between $a=10^{-4}$ and $a=10^{-3}$.

\subsection{Sound speed}\label{sec:sound_speed}

Describing the background expansion of the Universe through the Friedmann equation (see eq.~\ref{eq:Friedmann}) only requires knowing how the density components evolve with scale factor. However, as mentioned in section~\ref{sec:intro}, this is not the case of the perturbations, which in general also require the pressure perturbation and the anisotropic stress to be specified (we always neglect the anisotropic stress). The relation between the rest-frame pressure and density perturbations (respectively $\delta \overline{p}$ and $\delta \overline{\rho}$) can be parameterized through the sound speed parameter $\overline{c}_s^2$ with
\begin{align}
    \delta \overline{p}(a,k) = \overline{c}_s^2(a,k)\delta \overline{\rho}(a,k)\,.
\end{align}
The sound speed describes how fast perturbations propagate through the fluid and is generally a function of both the scale factor, 
$a$, and the scale, $k$, directly influencing the fluid's clustering behaviour (see also the perturbation equations~\ref{eq:perturbations1} and~\ref{eq:perturbations2}). Note that in the case where the equation of state parameter is $w=-1$ (i.e. a cosmological constant), the fluid has no perturbations and the sound speed parameter is meaningless. In this paper, we neglect the scale dependence of the sound speed and only probe its scale factor evolution between $a_1=10^{-5}$ and $a_2=10^{-3}$, which corresponds to the early Universe period in which we are interested. We therefore use the following simple parameterization,
\begin{align}
    \overline{c}_s^2(a) = 
        \begin{cases}
          c_1^2 & \text{if $a \leq a_1$} \\
          c_1^2 + (c_2^2-c_1^2)\frac{\log{a}-\log{a_1}}{\log{a_2}-\log{a_1}} & \text{if $a_1 \leq a \leq a_2$}\\
          c_2^2 & \text{if $a \geq a_2$} \,,
        \end{cases}   
\end{align}
which represents the sound speed of the EDF in its rest-frame and where $c_1^2$ and $c_2^2$ are the free parameters to be fitted to the observations within the range $0\leq c_1^2,c_2^2\leq 1$. Again, the idea here is not to accurately capture the exact behaviour in specific models, but to have a sufficiently general model that most new physical models will project at least one combination of the parameters that we vary into a value that differs from \lcdm.

To summarize, our model adds six new parameters to capture new physics in the early Universe. Four of them ($d_1, d_2, d_3, d_4$) are the amplitudes of the density modes used to model the density of the EDF, whereas the remaining two ($c_1^2$ and $c_2^2$) parameterize its sound speed as a function of scale factor. The EDF model is implemented as a modified version of CAMB that is publicly available\footnote{\href{https://github.com/raphkou/CAMB_EDF}{https://github.com/raphkou/CAMB\_EDF}}.

\section{Theoretical model tests}
\label{sec:test_models}
In order to test what kind of theoretical models our model can reproduce, we generate a set of smooth (theoretical) CMB temperature $T$, E polarization, and cross-correlation $TE$ power spectra for different models that we then fit with our model as though they were a data realization. We therefore effectively assess how accurately our model can reproduce the CMB power spectra for each theoretical model, rather than how well it directly reproduces the density or sound speed evolutions with scale factor. The models we test can be classified into EDE or dark radiation models and are presented in the following subsections. We also test the standard \lcdm\ case to assess any biases and parameter volume effects, and to forecast how well our model will be constrained by \SO\ observations in case that there is no deviation from \lcdm.

\subsection{Test cases}
\subsubsection{Axion-like EDE}\label{sec:axion_EDE}

The first model we consider is the so-called axion-like EDE~\cite{Poulin19,Smith20} which consists in the addition of a scalar field $\phi$ with a potential,
\begin{align}
    V(\phi)=m^2f^2\left[1-\cos(\phi/f)\right]^n=m^2f^2\left[1-\cos(\theta)\right]^n\,,
\end{align}
where $m$ and $f$ are the axion mass and decay constant, $n$ is an exponent, and $\theta=\phi/f$ is a re-normalized field variable comprised in $[-\pi,\pi]$. In the case where $n=1$, this potential is simply the axion potential that motivates this parameterization, whereas higher values of $n$ correspond to a more phenomenological approach that are necessary to potentially solve the Hubble tension~\cite{Poulin19}. As discussed in section~\ref{sec:intro}, this EDE model introduces a scalar field that is initially frozen but becomes dynamical during a specific phase of the Universe's expansion, eventually oscillating around the minimum of its potential. Once dynamical, the field dilutes like matter when $n=1$, like radiation when $n=2$ and faster than radiation for $n\geq 3$. Additional density that do not decay significantly faster than matter are strongly constrained by their effect on the early ISW effect~\cite{Vagnozzi21}.

The axion-like EDE can be modelled approximately as an effective-fluid with density,
\begin{align}
    \Omega_\phi(a) = \frac{2\Omega_\phi(a_c)}{(a/a_c)^{3(w_n+1)}+1}\,, 
\end{align}
with $w_n = (n-1)/(n+1)$ and $a_c$ being the scale factor at which the field becomes dynamical. In the effective-fluid approach, the sound speed of the fluid is $c_s^2=1$ when the field is frozen whereas when the field oscillates around the minimum of the potential, the sound speed~\cite{Poulin18} follows,
\begin{align}
    c_s^2(a,k) = \frac{2a^2(n-1)\varpi^2 + k^2}{2a^2(n+1)\varpi^2 + k^2}\,,
\end{align}
where $\varpi$ is the angular frequency of the oscillations (see~\cite{Poulin18} for more details). In this paper, we test the case $n=3$ which seemed to be the most promising regarding the Hubble tension. In particular, we use the best-fit parameters from~\cite{Poulin19}, i.e. $\log_{10}(a_c) = -3.696$ and $f_\textrm{EDE}(a_c)=0.058$ (with $f_\textrm{EDE}(a_c) = \Omega_\phi(a_c)/\Omega_\textrm{tot}(a_c)$). The CMB temperature and polarization power spectra are generated using the axion-like effective fluid implementation of CAMB (equivalent to the model of ref.~\cite{Poulin18}).

\subsubsection{New early dark energy}

The New Early Dark Energy (NEDE)~\cite{Niedermann21,Niedermann20} is another EDE model that is characterized~\cite{Cruz23} by a trigger mechanism inducing a sudden decay of an early dark energy component. Different mechanisms can be responsible for that sudden decay and lead to different NEDE models. In the Cold NEDE scenario that we study (and that is described in the references mentioned earlier), two scalar fields $\phi$ and $\psi$ are being considered. The subdominant field, $\phi$, is initially frozen because of Hubble friction while the other field $\psi$ cannot reach its minimum due to a potential barrier. As the Universe expands, $\phi$ starts to oscillate and approaches a threshold value at which the potential barrier is removed, allowing the two fields to transit to their global minimum. The NEDE energy density is defined as the potential energy released during the transition of the two fields and decays rapidly after $\phi$ reaches its threshold value and $\psi$ becomes dynamical.

As in the axion-like EDE case, the NEDE can be modelled using a fluid approach. In the NEDE case, the background equation of state of the fluid follows~\cite{Cruz23},
\begin{align}
    w_\textrm{NEDE}(t) = 
        \begin{cases}
          -1 & \text{if $t<t_*$} \\
          w_\textrm{NEDE}(t) & \text{if $t\geq t_*$}\,,
        \end{cases}   
\end{align}
where $t_*$ is the time at which $\phi$ crosses the threshold value and the equation of state is approximated as a constant after $t_*$ such that $w_\textrm{NEDE}(t>t_*)=w_\textrm{NEDE}^*$. Similarly, the sound speed is assumed to be a constant equal to the equation of state parameter, so $c_s^2=w_\textrm{NEDE}^*$. We generate a set of CMB temperature and polarization power spectra using the Boltzmann code TriggerCLASS~\cite{Niedermann20}, which is a modified version of CLASS~\cite{Blas11}. We used high precision settings when generating the power spectra with TrigerCLASS, and when fitting our EDF model implemented in CAMB, to make sure that what our model detects is not due to differences between CAMB and CLASS implementations. For a fiducial \lcdm\ model with the high precision settings we used, we obtained a $\chi^2$ as low as $0.34$ using \SO\ noise. As a result, the impact of differences between CAMB and CLASS implementations on our findings should be very limited. The fiducial value we use corresponds to the best-fit of the main analysis of ref.~\cite{Cruz23} which is $f_\textrm{NEDE}=0.12$, $3w_\textrm{NEDE}^*=1.902$ and $z_*=4290.739$ where $f_\textrm{NEDE}$ is the NEDE fraction at $t=t_*$ and $z_*$ is the redshift corresponding to $t_*$.

\subsubsection{Dark radiation \texorpdfstring{$N_\textrm{eff}$}{Neff}}

As introduced in section~\ref{sec:intro}, one way of reducing the Hubble tension is through the addition of relativistic, non-interacting species that constitute a dark radiation and are usually parameterized through equation~\ref{eq:Neff} that we recall here for convenience:
\begin{equation}
    \rho_R=\rho_\gamma\left(1+\frac{7}{8}\left(\frac{4}{11}\right)^{4/3}N_\textrm{eff}\right)\,. \tag{\ref{eq:Neff} repeated}
\end{equation}
Here, $N_\textrm{eff}$ is the effective number of relativistic species equal to $N_\textrm{eff}=3.044$ in the Standard Model of particle physics, taking into account neutrino heating by electron-positron
annihilation~\cite{Hannestad95,Dolgov97,Gnedin98,Mangano05,DeSalas16}. Additional relativistic particles would contribute to the radiation density; for a given model, this contribution to $N_\textrm{eff}$ can be theoretically computed. In a phenomenological approach, it has however become usual to directly make $N_\textrm{eff}$ a free parameter in cosmological analyses. Particles that contribute to $N_\textrm{eff}$ are usually considered to be non-interacting (like neutrinos),  and in particular have nonzero anisotropic stress, which in general requires computation of the  full Boltzmann hierarchy (though absence of self-interactions is not assumed in the background-level, Eq.~\ref{eq:Neff}). In this paper, we make use of the $N_\textrm{eff}$ implementation in CAMB to generate the fiducial CMB power spectra, taking $N_\textrm{eff}=3.33$ which corresponds to the $95\%$ upper limit from ref.~\cite{Planck18}.

\subsubsection{Self-Interacting Dark Radiation (SIDR)}

Dark radiation particles could have different properties from neutrinos. One possibility is self-interacting dark radiation (SIDR)~\cite{Jeong13,Cyr16,Lesgourgues16,Khalife24}. Frequent scatterings between dark radiation particles would prevent free-streaming, and suppress all multipole moments of the distribution function with $\ell \geq 2$. In the Generalized Dark Matter formalism, such a self-interacting dark radiation would hence correspond to an equation of state parameter and sound speed equal to $w=c_s^2=1/3$, and no anisotropic stress. We implement this self-interacting dark radiation in CAMB, choosing a density at $z=0$ that matches the previous case $N_\textrm{eff}=3.33$.

\subsection{Results}
For all test cases presented earlier, we use COBAYA \cite{Torrado19,Torrado21} to conduct Monte Carlo Markov Chain (MCMC)~\cite{Lewis:2002ah,Lewis:2013hha} analyses. We use a Gaussian likelihood with the same covariance matrix as that used to estimate the best-constrained modes through the Fisher formalism. Specifically, for each fiducial model we generate TT, EE, and TE theory spectra and use a covariance appropriate for \SO\ over 40\% of the sky, using multipoles in the range $40 < \ell < 5000$.
We use flat priors in the ranges $[0,0.15]$ and $[0,1]$ for our four density parameters and two sound speed parameters, respectively. We also use flat priors for the \lcdm\ parameters, except $\tau$ for which we add a Gaussian prior with mean $\mu=0.0561$ and standard deviation $\sigma=0.0073$. This prior replaces the constraint that would be obtained by adding the low multipoles from \Planck. All the chains are converged with the Gelman-Rubin criterion $R-1<0.05$. Marginalized parameter results are obtained from the Monte Carlo samples using GetDist~\cite{Lewis:2019xzd}.

We also ran chains for future observations matching \lcdm , giving us a reference to compare $\chi^2$ values, a way to assess the improvement in fit when using our model compared to \lcdm, and to show the parameter volume/degeneracy effects that will always be present in marginalized results. Because our model has $6$ additional parameters compared to \lcdm, comparing the \(\chi^2\) values alone may not always be a fair criterion, so we also use the Akaike Information Criterion (AIC)~\cite{Akaike74,Trotta17}, defined as:
\begin{align}
    \textrm{AIC} = 2k-2\log{\mathcal{L}},
\end{align}
where $\mathcal{L}$ is the best-fit likelihood and $k$ is the number of parameters in the model ($6$ for \lcdm\ and $12$ for EDF). We then computed the AIC difference in the two models such that,
\begin{align}
    \Delta\textrm{AIC}=\textrm{AIC}(\Lambda\textrm{CDM})-\textrm{AIC}(\textrm{EDF}).
\end{align}
The EDF model would be disfavoured compared to \lcdm\ if $\Delta\textrm{AIC}<0$, would perform as well as \lcdm\ if $0<\Delta\textrm{AIC}<2$ and would be favoured otherwise.
\subsubsection{\texorpdfstring{\lcdm}{LCDM} case}

\begin{figure}[htbp]
\centering
\includegraphics[width=\textwidth]{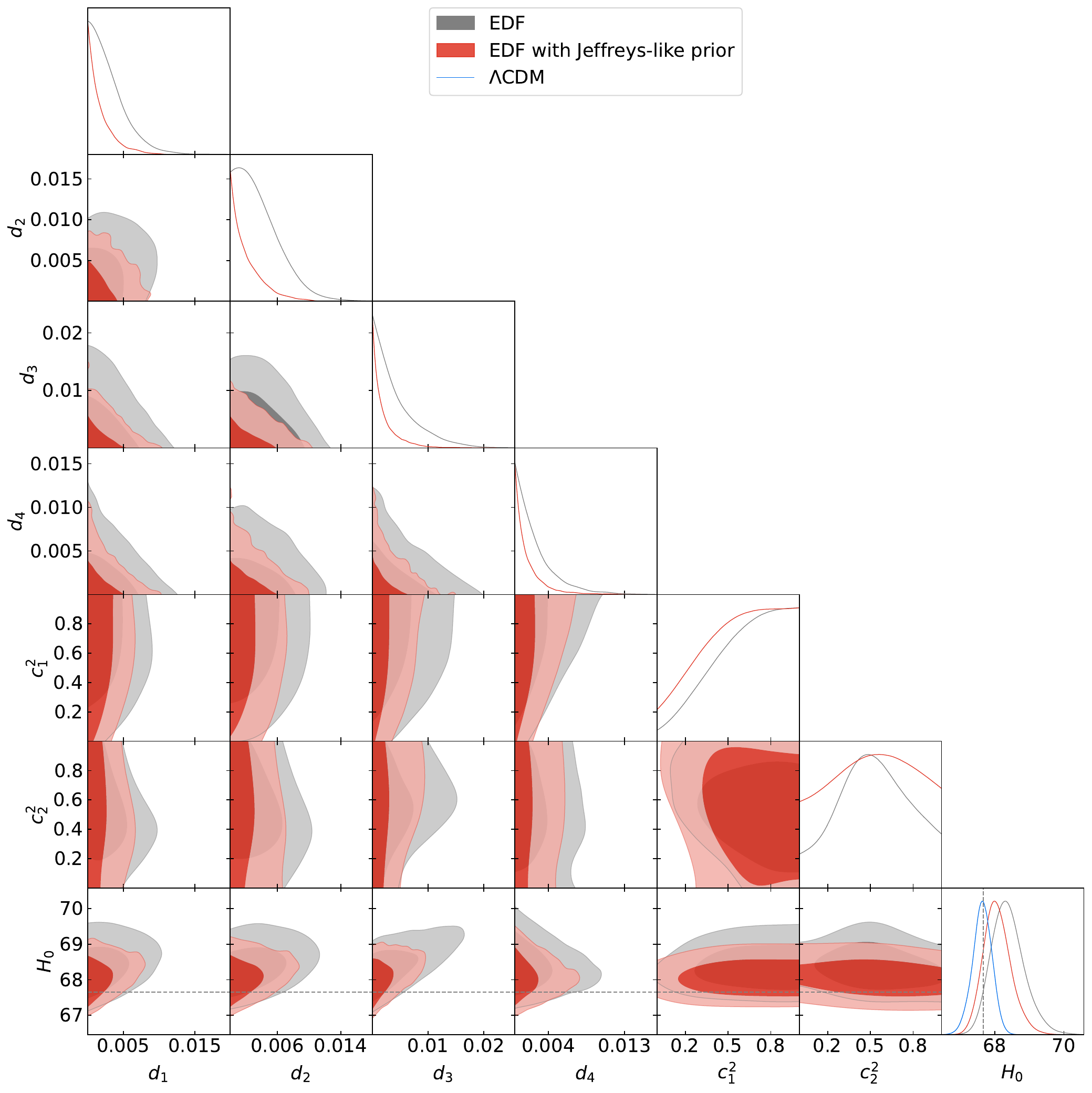}
\caption{Forecast constraints on the six EDF parameters and $H_0$ when fitting CMB power spectra generated with a \lcdm\ underlying fiducial model using noise curves for \SO\ (grey contours). The blue posterior for $H_0$ is obtained when using \lcdm\ to fit the power spectra and serves as a reference. We additionally show the fiducial value of $H_0$ as a grey dashed line. The red contours show results using a non-flat prior on the density parameters, using a regularization parameter $\alpha=10^{-3}$ (see section~\ref{sec:jeffrey}).\label{fig:triangle_plot_lcdm} Contours contain $68\%$ and $95\%$ of the marginalized probability.}
\end{figure}

We first look at \SO\ constraints in the case of a fiducial \lcdm\ model. Forecast constraints on the fluid parameters and $H_0$ are shown in figure~\ref{fig:triangle_plot_lcdm} (grey). All the density amplitudes are consistent with $0$, showing no detection of deviation from \lcdm. The sound speed parameters are unconstrained, which is expected since those parameters are meaningless in the \lcdm\ case. However, their posterior is not completely flat and shows a preference for higher values. This is a volume effect coming from the fact that larger additional density is allowed if it does not cluster too strongly, i.e. the sound speed is significantly greater than zero. Larger additional densities are implicitly preferred by having a  positivity prior on multiple independent redshift bins.
Intriguingly, we also observe that simultaneous low values of $c_1^2$ and $c_2^2$ seem to be excluded (see the $c_1^2$--$c_2^2$ plane of figure~\ref{fig:triangle_plot_lcdm}). This is also due to a volume effect, as all sound speed values should be allowed when all density amplitudes are very close to zero. However, the volume where all the new parameters are close to zero only occupies a very small fraction of the high-dimensional posterior parameter space. A more in-depth demonstration of how sound speed parameters affect the power spectra can be seen in figure~\ref{fig:cl_ee_cs2}.

Finally, we observe that the posterior of $H_0$ is slightly ($1.49\sigma$) skewed towards higher values compared to the posterior we get using the \lcdm\ model (blue on figure~\ref{fig:triangle_plot_lcdm}). This comes from the fact that we only allow for positive density amplitudes, which always lead to higher values of $H_0$ at a given angular acoustic scale $\theta_*$ (which is almost fixed by the data). However, this effect is small for \SO\ and would not result in a false inference of a value of $H_0$ high enough to resolve the Hubble tension (it could nevertheless reduce slightly the tension, so that the results would need to be analysed with great care). We could however also use a prior on the density amplitudes to reduce this effect further (red on figure~\ref{fig:triangle_plot_lcdm}), as we discuss in more detail in section~\ref{sec:jeffrey}. We show the correlations between all the parameters (including the six \lcdm\ parameters) in appendix~\ref{app:triangle_plot_lcdm_full_parameters} (figure~\ref{fig:triangle_plot_lcdm_full_parameters}). It can be seen that the fiducial parameters used to generate the spectra are always recovered within $1\sigma$, except $\Omega_ch^2$ which is slightly shifted towards higher values (at $1.62\sigma$ from the fiducial value).

\subsubsection{EDE models}

\begin{figure}[htbp]
\centering
\includegraphics[width=\textwidth]{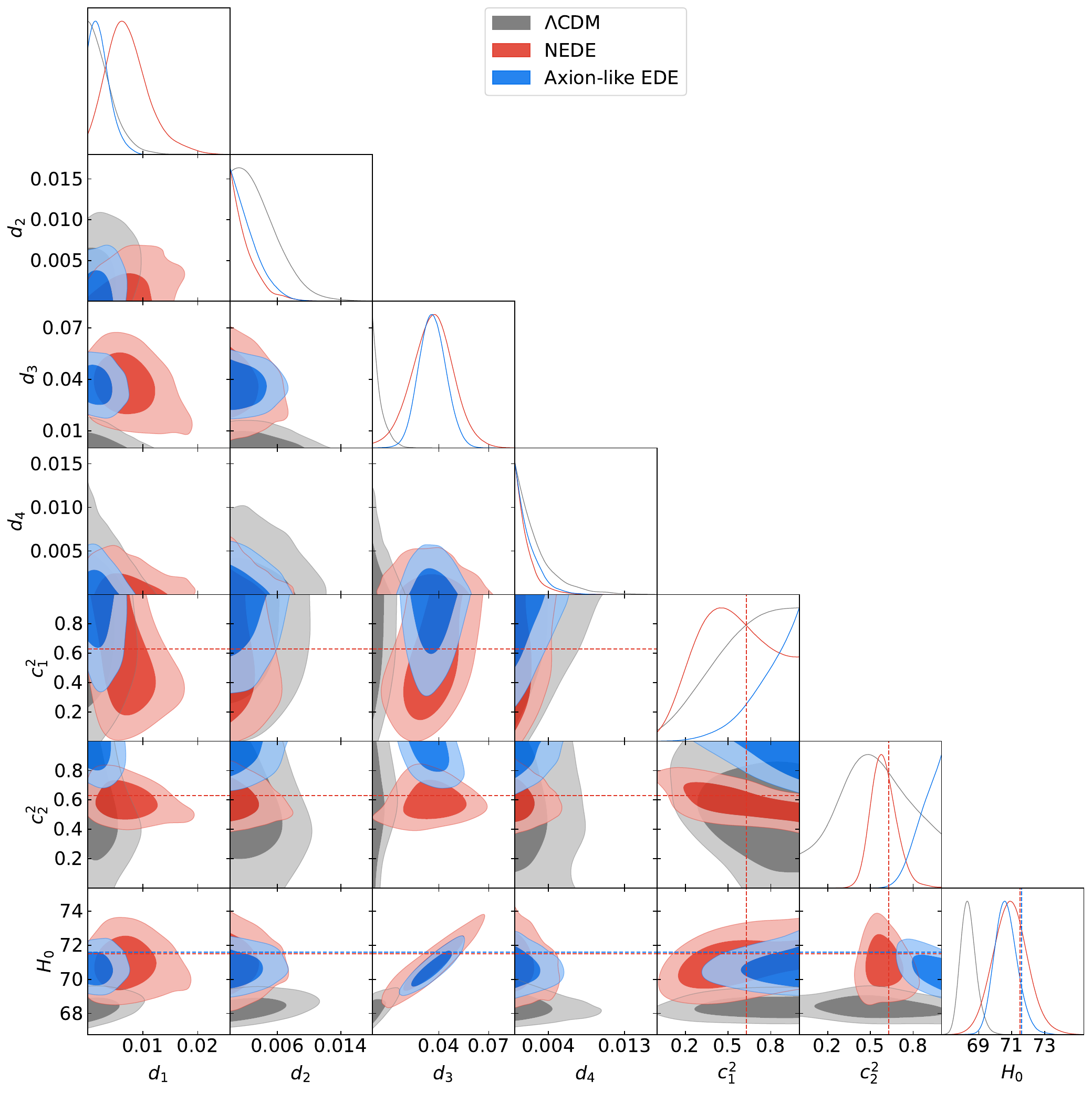}
\caption{Forecast constraints on the six EDF parameters and $H_0$ when fitting CMB power spectra generated with a \lcdm\ (grey), NEDE (red) or Axion-like EDE (blue) underlying fiducial using the noise curves for \SO. The blue and red dashed lines show the fiducial value of $H_0$ for the two test cases. We also show the sound speed value used in the fiducial model in the NEDE case with red dashed lines.\label{fig:triangle_plot_EDE}}
\end{figure}

We show the constraints we would be able to place on the two EDE models (assuming our test parameters are the truth) using \SO\ data on figure~\ref{fig:triangle_plot_EDE}. We also show the \lcdm\ constraints in grey for reference (in the case with flat priors). The two EDE models trigger the same density amplitude $d_3$, but the NEDE model also prefers a small contribution from \(d_1\), unlike the axion-like EDE model. The other density amplitudes do not contribute significantly to reproducing these models. It is particularly interesting to see that these two EDE models are reproduced very similarly, indicating that our parameter \(d_3\) is able to capture a broad range of EDE models. This parameter is highly correlated with \(H_0\), which is expected since it captures the most relevant redshift range for the effect of the EDE models on the CMB. In addition, the posteriors of $H_0$ are in good agreement with the test value used for the two EDE models. The fiducial sound speed value used in the NEDE, which was $c_s^2=0.63$ (shown in dashed line on figure~\ref{fig:triangle_plot_EDE}), is very well recovered by $c_2^2$ (value of the sound speed at $a=10^{-3}$) whereas $c_1^2$ is not very well constrained but is still compatible with the fiducial value. In the axion-like EDE case, the sound speed in general depends on both scale and scale factor, so it is more difficult to interpret the posteriors we obtained. However, in the axion-like EDE model the sound speed tends to $1$ as $n$ increases, so for $n=3$, having posteriors compatible with $c_s^2=1$ is consistent. Other parameters constraints and correlations can be found in appendix~\ref{app:triangle_plot_lcdm_full_parameters} (figure~\ref{fig:triangle_plot_EDE_full_parameters}) where it can be seen that most parameters are well recovered apart from few biases on~$\theta_*$ and $n_s$.

\begin{table}[htbp]
\centering
\begin{tabular}{cccccc}
\hline
Theoretical model & $\chi^2_\textrm{EDF} $ & $\chi^2_{\Lambda\textrm{CDM}}$ & $\Delta \chi^2$ & $\Delta \textrm{AIC}$ & $f_\textrm{max}$\\
\hline
Axion-like EDE & $4.58$ & $30.13$ & $-25.55$ & $13.55$ & $0.068$ \\
NEDE & $10.33$ & $36.72$ & $-26.39$ & $14.39$ & $0.120$ \\
\Neff & $2.09$ & $16.29$ & $-14.20$ & $2.2$ & $0.037$ \\
SIDR & $0.81$ & $20.08$ & $-19.27$ & $7.27$ & $0.037$ \\
\hline
\end{tabular}
\caption{Comparison of how well our model can reproduce a variety of theoretical models compared to \lcdm. Note that we are using all multipoles in the range $40\leq\ell\leq5000$ of the TT, EE and TE power spectra. If one of the two EDE models we tested were the truth, our model would be able to detect it compared to \lcdm\ with a strong evidence. In the dark radiation cases, the evidence is not that strong (especially when looking at the $\Delta\textrm{AIC}$) but this merely comes from the fact this criterion penalizes the additional parameters quite strongly, since the absolute $\chi^2$ of our model is excellent, showing that it can capture most of the effect of those models at the level of the CMB power spectra. We also show $f_\textrm{max}$, the maximum density fraction of the added component (EDE or dark radiation) in the fiducial test cases. It can be seen that the $\chi^2$ values are quite sensitive to these fractions, as they are a way of quantifying the deviations from \lcdm.\label{tab:chi2}}
\end{table}

Finally, we show the $\chi^2$ values of the best-fit samples obtained for the two models in table~\ref{tab:chi2}. Since we are fitting noiseless power spectra, the $\chi^2$ would be $0$ if our model was able to exactly reproduce the power spectra. The second column of the table also shows the minimum $\chi^2$ value we obtain when fitting the standard \lcdm\ model to the spectra. In the axion-like EDE and NEDE cases, the $\chi^2$ is reduced by approximately $25.55$ and $26.39$, leading to a $\Delta\textrm{AIC}$ of $13.55$ and $14.39$ respectively. Those values show that our model would be detected with strong evidence in the two cases. 

In our EDF model, the first and third modes peak at $\log_{10}(a)=-4$ and $\log_{10}(a)=-3.6$ respectively. As mentioned in section~\ref{sec:axion_EDE}, the fiducial parameters we used for the axion-like EDE are the best-fit values from ref.~\cite{Poulin19}, which in particular include $\log_{10}(a_c)=-3.696$. Because this value is rather close to the peak of our third mode, we checked whether that was the reason why our model could detect deviations from \lcdm\ with high significance. To do so, we changed the fiducial value to $\log_{10}(a_c)=-3.8$, in the middle of our first and third modes. Using the EDF model, we obtained a best-fit sample $\chi^2=12.51$, showing that indeed our model could not reproduce it as well as when $\log_{10}(a_c)=-3.696$. However, fitting the same power spectra with the \lcdm\ model, we obtained a best-fit sample $\chi^2=51.23$. Choosing $\log_{10}(a_c)=-3.8$ therefore leads to very strong deviations from \lcdm\ and the EDF model can reduce the $\chi^2$ very efficiently. As a result, the $\Delta\textrm{AIC}$ value would be $26.72$ so the EDF would be detected with even higher evidence. In conclusion, our model seems to be able to reproduce the EDE models we tested quite well at the power spectra level and the parameters (especially $H_0$ and the sound speed parameters) are in good agreement with the underlying theoretical models.

\begin{figure}[htbp]
\centering
\includegraphics[width=\textwidth]{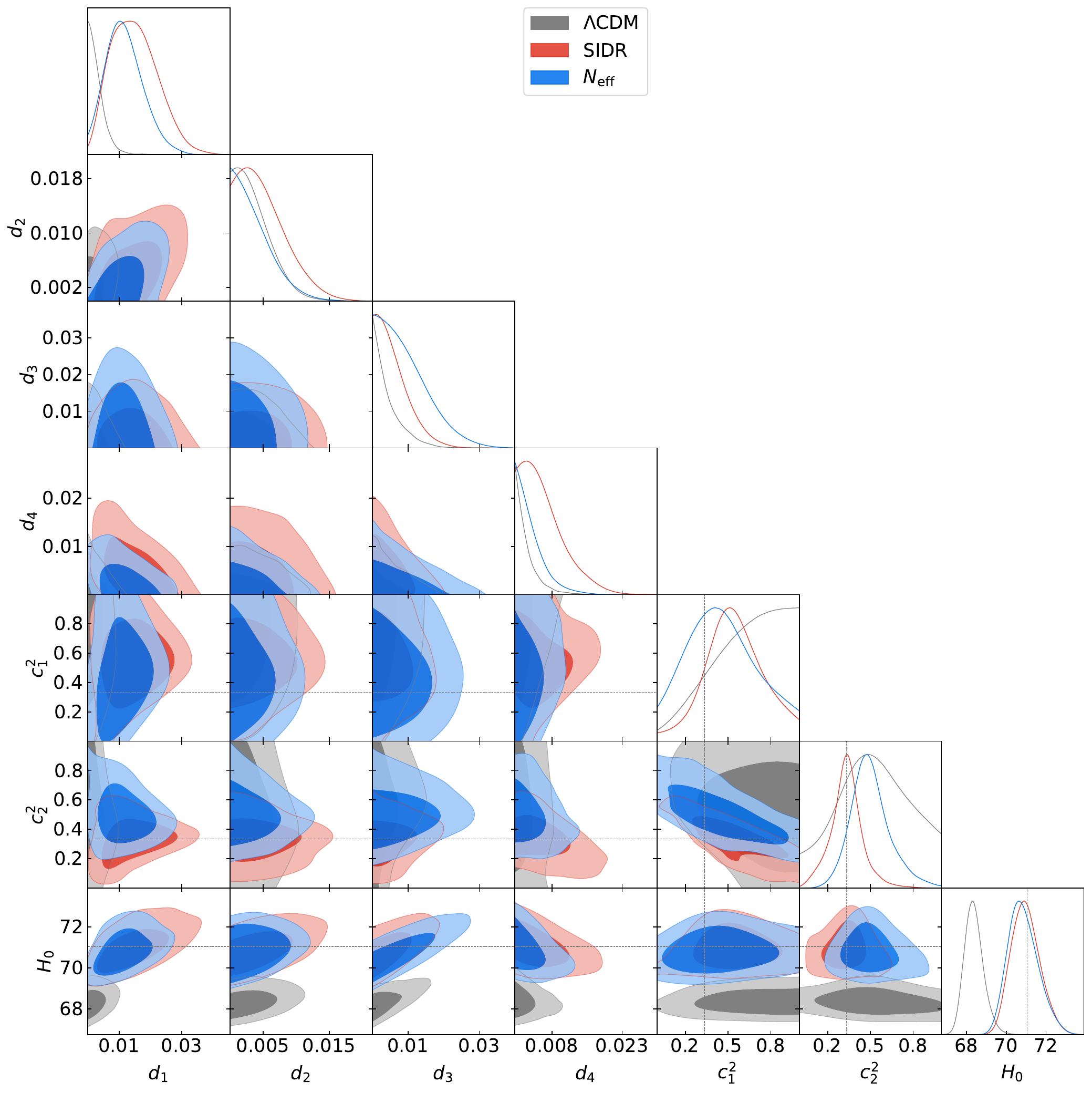}
\caption{Forecast constraints on the six EDF parameters and $H_0$ when fitting CMB power spectra generated with a \lcdm\ (grey), SIDR (red) or \Neff\ (blue) underlying fiducial using noise curves of \SO. The grey dashed lines show the sound speed (i.e. $c_s^2=1/3$) and $H_0$ values used in the fiducial model in the SIDR and \Neff\ cases.\label{fig:triangle_plot_DR}}
\end{figure}

\subsubsection{Dark radiation models}

As for the EDE models, figure~\ref{fig:triangle_plot_DR} shows the forecast constraints on dark radiation models. The \Neff\ case is shown in the blue contours, and the SIDR model in red. Finally, we show the \lcdm\ constraints for comparison in grey. This time, it appears that the dark radiation models are mainly reproduced through the $d_1$ parameter, even though $d_3$ can also take values of the order of a few percent. The two models are reproduced very similarly, which is not surprising given that they only differ by their anisotropic stress and higher multipole moments, which our model does not include. As for the EDE models, the central value of the posterior of $H_0$ is in excellent agreement with the value predicted with the underlying model. Figure~\ref{fig:triangle_plot_DR} shows that $H_0$ is highly correlated with $d_1$ and $d_3$, the two parameters that seem to be describing the dark radiation density. Figure~\ref{fig:triangle_plot_DR_full_parameters} from appendix~\ref{app:triangle_plot_lcdm_full_parameters} shows the constraints and correlations between the $12$ (six \lcdm\ and six EDF) parameters varied in this analysis. Apart from a slight bias on $A_se^{-2\tau}$, all parameters are well recovered.

The expected sound speed parameters for dark radiation are $c_1^2=c_2^2=1/3$, shown in grey dashed lines in the figure. The first parameter $c_1^2$ is unconstrained, but the posterior of the second parameter $c_2^2$ agrees well with the expected value in the SIDR case. In the \Neff\ case, the posterior peaks at a slightly higher value, but still has a quite high probability at $c_2^2=1/3$. The slight shift towards higher values could be due to the fact that our model does not include any free-streaming (anisotropic stress and higher-multipole) effects, unlike \Neff, so the free-streaming phase shift in the perturbations due to the neutrinos (see~\cite{Bashinsky04,Baumann16}) may be being partially mimicked by our (non-free-streaming) fluid having a sound speed $c_2^2>1/3$ (best-fit sample having $c_2^2=0.49$), larger than in the SIDR case. Table~\ref{tab:chi2} shows that our model is able to reproduce the CMB power spectra of these models with high accuracy, with the best-fit sample $\chi^2$ being 0.81 and 2.09 for SIDR and \Neff\ respectively. The higher $\chi^2$ for \Neff\ is expected since our model does not include any anisotropic stress, but the $\chi^2$ is still relatively low. However, the $\Delta\textrm{AIC}$ is not very large, especially in the \Neff\ case. This is because the $\chi^2$ value for the \lcdm\ model is already reasonably good for these models, and the AIC heavily penalizes the introduction of additional parameters. This is however inevitable with our approach since we aim at reproducing a large variety of models, which hence requires introducing a significant number of new parameters. 
 Well-motivated physical models that just contribute additional radiation that give a pure \Neff\ effect would of course be tested separately as a one-parameter \lcdm\ model extensions, but our model can potentially also probe more complicated models with other contributions.
 Overall, our model is able to reproduce the CMB observations of the two dark radiation models with very high accuracy and, as for the EDE models, the values inferred for $H_0$ and the sound speed are consistent with the underlying fiducial values. 

\subsubsection{Discussion}

We showed on the four test cases (axion-like EDE, NEDE, \Neff\ and SIDR) that our model reproduces the CMB power spectra predicted by those models much better than \lcdm. In addition, the inferred value of $H_0$ as well as the sound speed parameters estimated with our model are consistent with the fiducial models used to generate the power spectra. This demonstrates the ability of our model not only to detect deviations from \lcdm, but also to provide clues about the type of new physics in the early Universe. The results based on the four test cases show that our model is able to discriminate broad classes of models, as EDE or dark radiation models lie in different regions of the parameter space. However, the fact that some different EDE or dark radiation models project very similarly into our model parameters means that it will remain hard to distinguish between some models. Additionally, we did not consider models with new interactions between the fluids, but it is plausible that many such models would also project significantly onto our parameterization, allowing deviations from \lcdm\ to be detected.

\begin{figure}[htbp]
\centering
\includegraphics[width=\textwidth]{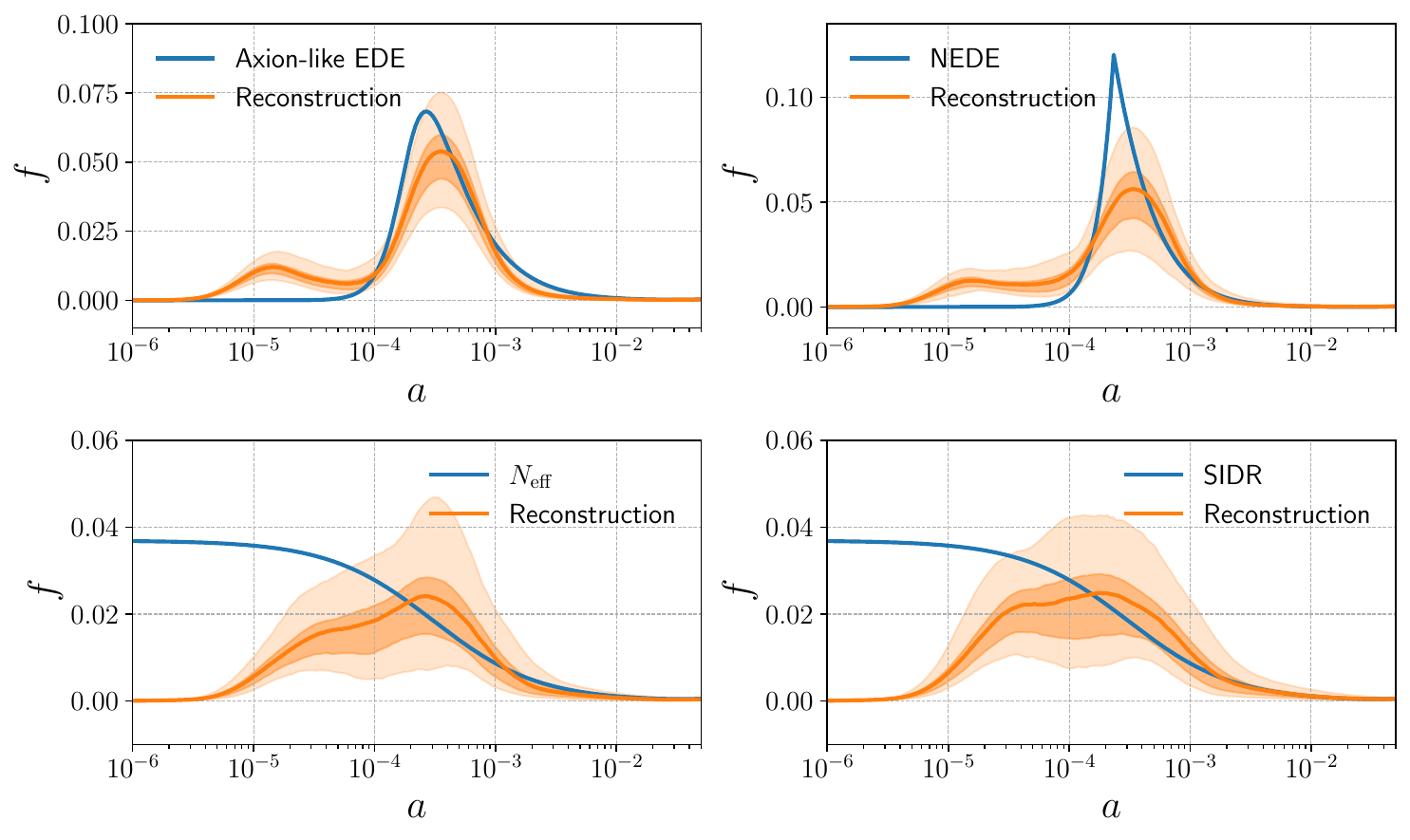}
\caption{Evolution of the fractional density of the added component $f$ as a function of scale factor. Blue curves show the theoretical evolution of the fractional density of the added component (either early dark energy or additional radiation), while orange curves show the median of the EDF reconstruction (obtained after having fitted the data). Shaded areas represent the $68\%$ and $95\%$ confidence intervals.\label{fig:reconstructed_density}}
\end{figure}

\begin{figure}[htbp]
\centering
\includegraphics[width=0.7\textwidth]{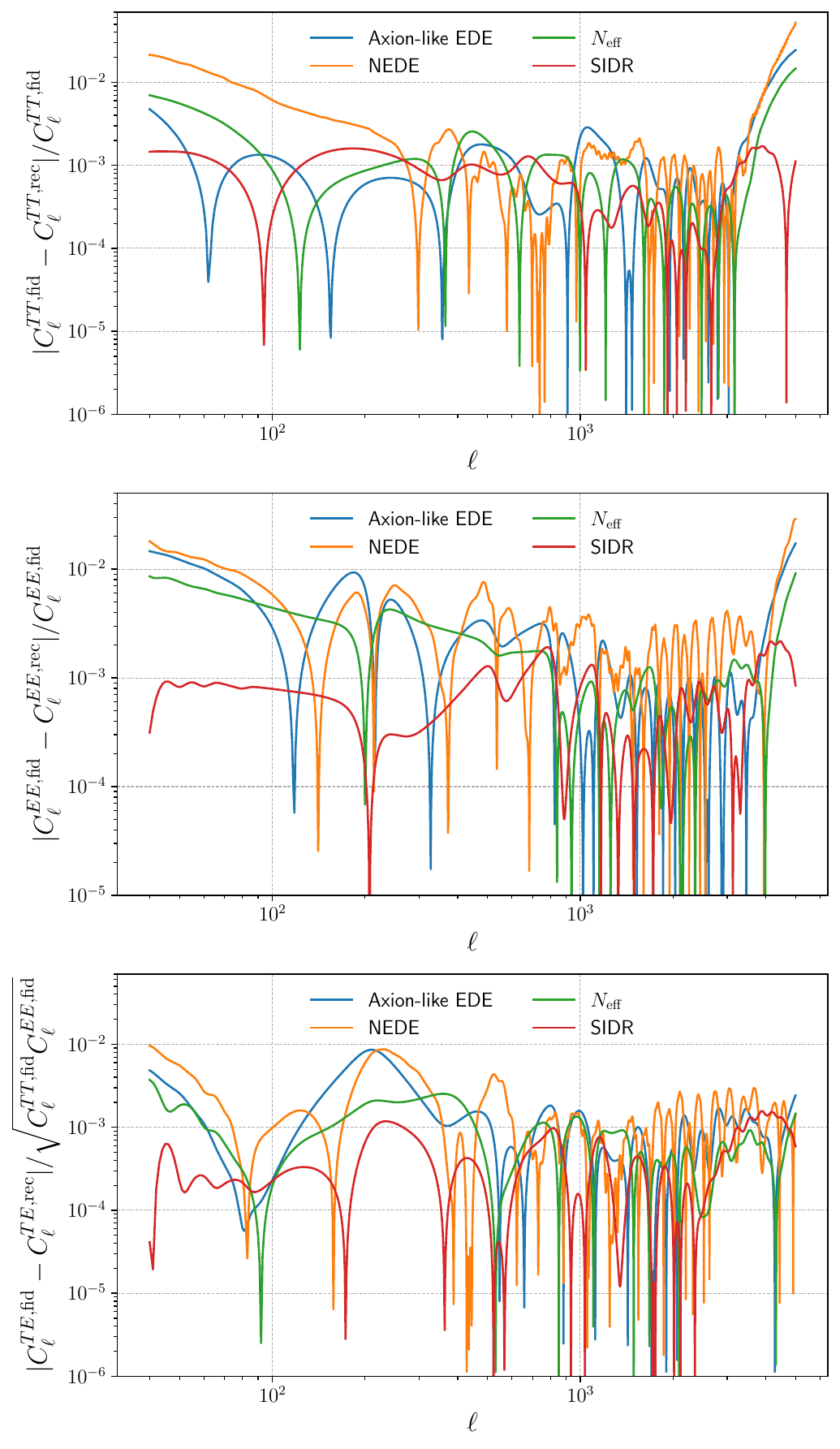}
\caption{Relative differences between the theoretical power spectra and the best-fits of our model. For the three power spectra and the four models, relative differences are mostly well below the percent-level, except at very small and very large scales.\label{fig:power_spectra}}
\end{figure}

Our model's main performance measure is the fit to the CMB power spectra. However, we can also compare how the added component's density changes with the scale factor in both the original theoretical model and our reconstructed fit. Figure~\ref{fig:reconstructed_density} shows this comparison for four test cases. Blue lines show the added component's fraction of total density in theoretical models. Orange lines show the median of this fraction in our reconstructed model posterior. Shaded areas mark the 68\% and 95\% confidence ranges.
Additionally, figure~\ref{fig:power_spectra} shows the relative differences between the CMB power spectra predicted by the fiducial models and the best-fit spectra in our model. As we saw in the parameter constraints, figure~\ref{fig:reconstructed_density} shows that the two EDE models are reconstructed similarly by the EDF model. This is also true of the two dark radiation models. The axion-like EDE model is generally well reproduced, except for a small peak around $a=10^{-5}$. Our model reconstructs the NEDE comparably well, but cannot reproduce the sharp peak of the NEDE density because of the underlying smooth spikes that we are using. For the dark radiation models, the fractional density is well reproduced for scale factors greater than $a>10^{-4}$ where the fiducial model mostly lies in our $68\%$ confidence interval. However, the reconstruction is less accurate at smaller scale factors, where the impact on the CMB power spectra is smaller.

Figure~\ref{fig:power_spectra} shows that in all cases, our model reproduces the power spectra to much better than percent-level accuracy for most multipoles. This is especially the case in the range $\ell\in[100,3000]$ where \SO\ has the tightest constraints. Outside this range, where cosmic variance or noise are more important, there are larger differences. As indicated by the $\chi^2$ values of table~\ref{tab:chi2}, the power spectra fit better in the dark radiation cases.

As mentioned in section~\ref{sec:intro}, neither EDE nor dark radiation models are actually favoured by current CMB data as a solution to the Hubble tension. The deviations from \lcdm\ power spectra, in the region of parameter space where the models could help with the Hubble tension, are mainly captured by the $d_1$ and $d_3$ parameters in our approach. The other two density parameters $d_2$ and $d_4$, in combination or not with $d_1$ and $d_3$, could hence be interesting ways of representing as-yet-unknown models and uncover hints of physics that current models do not explain, and more generally allow physical models more freedom to fit the data accurately. The parameter $d_3$ is likely of most interest for the Hubble tension as it peaks around matter-radiation equality, is highly correlated with $H_0$, and appears in our fit of both EDE and dark radiation models.

Because of the relatively large number of parameters involved in our model, it is unlikely to detect deviations from \lcdm\ with high significance in a single parameter. We suggest that the comparison between our model and \lcdm\ should be made on the basis of the $\chi^2$ and $\Delta\textrm{AIC}$ values, or more sophisticated methods including Bayesian evidence computations~\cite{Skilling04,Handley15,Handley15b,Jia19,Srinivasan24,Polanska24}. When testing our model with \SO\ data, finding a high value of $H_0$ without using any local measurement of $H_0$, and with a significant $\chi^2$ improvement, would be good evidence for a new physics solution to the Hubble tension.

\section{Constraints with \Planck\ data}
\label{sec:Planck_constraints}
\subsection{Results}

While this model was primarily designed to be tested with \SO, it is still valuable to look at the state of the art with existing data. We therefore run a MCMC using the \Planck\ NPIPE likelihood~\cite{Rosenberg22} for the high-$\ell$ TT, TE and EE power spectra, and the low-$\ell$ TT and EE 2018 likelihoods~\cite{Planck20}. We still use flat priors on our $6$ additional parameters as well as all the \lcdm\ parameters. We removed the Gaussian prior on $\tau$ since we included the low-multipole likelihoods.

\begin{figure}[htbp]
\centering
\includegraphics[width=\textwidth]{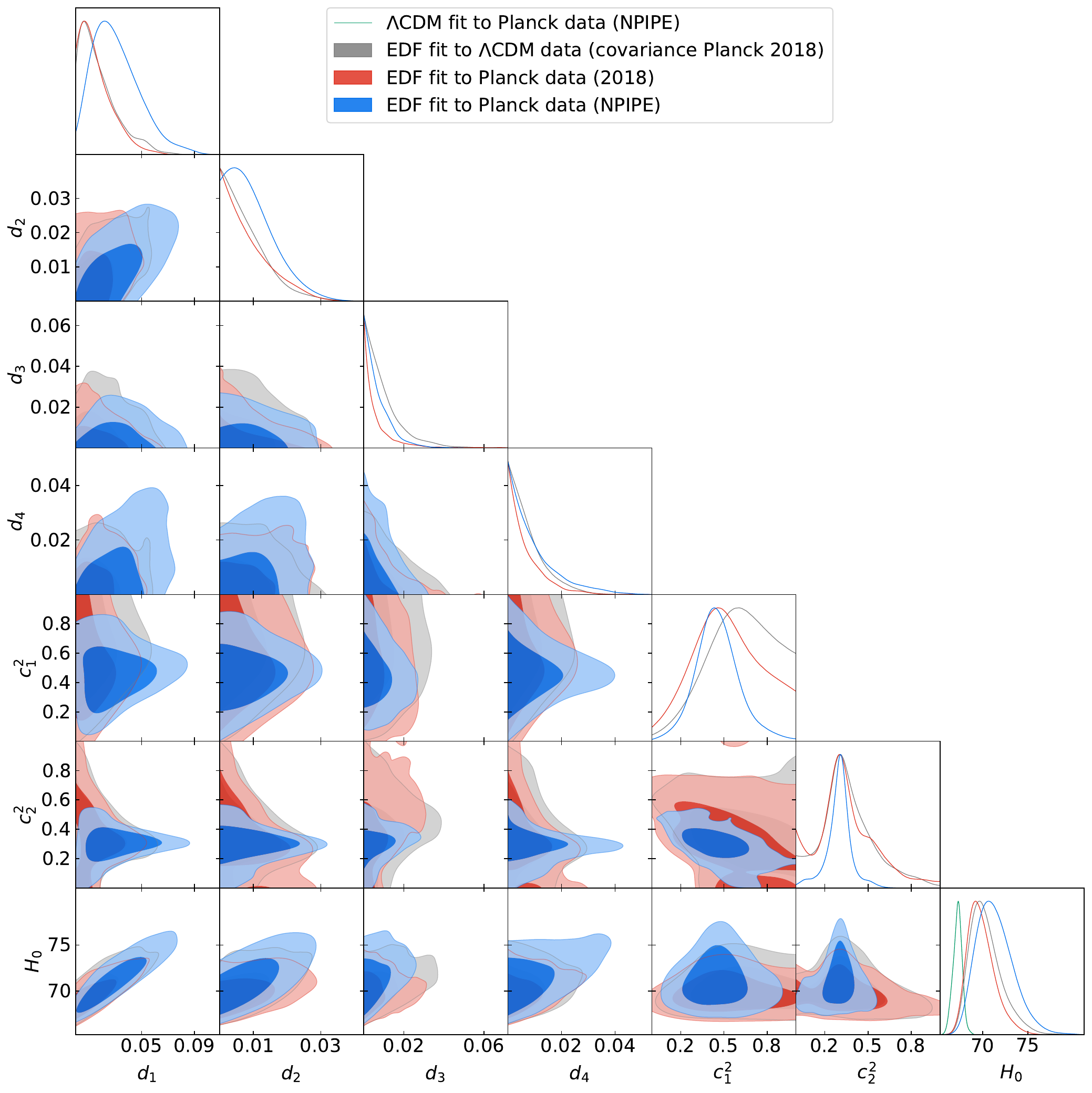}
\caption{Constraints on the six EDF parameters and $H_0$ when fitting our model to \Planck\ data (blue for the NPIPE likelihood, red for the 2018 lite likelihood) or to \lcdm\ theory 
power spectra with \Planck\ 2018 lite noise (grey). We also show the posterior of $H_0$ when fitting the \lcdm\ model to \Planck\ data (green).\label{fig:triangle_plot_Planck}}
\end{figure}

Figure~\ref{fig:triangle_plot_Planck} shows the constraints we obtained for the six EDF parameters and $H_0$, using \Planck\ data (blue contours). These are compared with the $H_0$ posterior obtained from a \lcdm\ model chain (green). Additionally, we provide the $95\%$ confidence interval and compare $\chi^2$ values in table~\ref{tab:Planck}. In the density parameter posteriors, we see a slight preference for a higher $d_1$, although this parameter remains consistent with zero. The other density modes are also consistent with zero, and constrained to within a few percent. 

When looking at the sound speed parameters, it looks as though the two parameters, especially $c_2^2$, are well constrained. This is however largely a volume effect, similar (yet much more pronounced) to what we forecast for \SO\ in the \lcdm\ case (see grey contours in figure~\ref{fig:triangle_plot_lcdm}). Indeed, we also ran a \lcdm\ chain to fit \Planck\ data (green posterior) and then used the best-fit values to generate a set of power spectra using the \lcdm\ prediction from CAMB. We then ran the EDF model on that mock \lcdm\ data using the \Planck\ 2018 lite likelihood~\cite{Planck20} for the high multipoles, where we replaced the observed data by our mock \lcdm\ data. We made use of the 2018 lite likelihood as it was easier to replace the data than in the NPIPE likelihood. As for the low multipoles, we used the actual \Planck\ observations for the low TT data, while we replaced the EE data by a Gaussian prior on $\tau$, having a mean $\mu=0.0561$ and standard deviation $\sigma=0.0073$ as we did previously for \SO. The grey contours in figure~\ref{fig:triangle_plot_Planck} show that while all density modes remain consistent with zero, there is apparently a strong marginalized preference for a high $c_1^2$ and $c_2^2 \sim 0.3$.  This stems from a volume effect: the \lcdm\ scenario (where all density modes are simultaneously zero) occupies a small portion of the posterior volume. Apart from the $(d_1, d_2, d_3, d_4) = (0, 0, 0, 0)$ configuration, a high $c_1^2$ and $c_2^2 \approx 0.3$ seems to allow larger values of the densities. 

Going back to the actual measurements, we observe the same effect with even slightly tighter constraints on the sound speed parameters. This probably comes from the preference for a high $d_1$. These results are therefore compatible with the \lcdm\ model, but also allow for a high $d_1$ with a sound speed of the order of $1/3$, which resembles the SIDR model studied earlier. This result agrees with ref.~\cite{Moss21}, who also found hints of an additional dark radiation. However, when looking at the $\chi^2$ of the best-fit sample (see table~\ref{tab:Planck}), the improvement is very slight (it only decreases by $3.7$) and the EDF model is even disfavoured when looking at the $\Delta\textrm{AIC}$. This occurs due to the large number of parameters added, which offer only a small improvement in $\chi^2$. Thus, while our model does not outperform 
\lcdm\ on Planck data, it is not ruled out either. The data still permit a high $d_1$, with an upper limit on most other density parameters around $0.02 - 0.03$ at $95\%$ confidence. When additional EDF density is introduced, the power spectra are more damped at small scales for fixed $\ell$. While this can be compensated by larger foreground amplitudes in Planck data, high-$\ell$ temperature measurements from ACT and SPT, along with future precision polarization data, will make these effects much more distinctive.

We also ran our EDF model using the \Planck\ 2018 lite likelihood, and obtained the red contours on figure~\ref{fig:triangle_plot_Planck}. This way, we can make a fair comparison between the constraints we get with the \lcdm\ mock data and the actual observations, using the same likelihood and error model. The contours we obtain are very close to the \lcdm\ case, demonstrating once again the strong consistency between \Planck\ data and the \lcdm\ model. The main differences when comparing the constraints obtained using either the NPIPE or the 2018 lite likelihoods are to be seen on the $d_1$ and the sound speed parameters: there is no preference for a high $d_1$ when using the 2018 lite likelihood, and the upper limit is tighter. The constraints on the sound speed parameters are also broader with the 2018 lite likelihood, likely because $d_1$ tends to be closer to zero more often. This could also be due to the NPIPE data excluding the larger phase shifts associated with high sound speeds (see section~\ref{sec:phase_shift}).

When looking at the posterior of $H_0$, we see it is centred around $71.3$ (using NPIPE), consistent with local measurements of $H_0$. However, this is also a volume effect, as previously observed with \SO\ noise curves: we are only allowing for positive density parameters, which always leads to the inference of high $H_0$ for a fixed angular scale $\theta_*$, and hence shifts the posterior of $H_0$ towards higher values. When looking at the posterior of $H_0$ using the EDF model with the \lcdm\ data, the same effect is to be seen: the posterior of $H_0$ is shifted towards larger values. However, although the $H_0$ posterior is centred around large values due to volume effects, it does clearly demonstrate that there are degenerate directions in parameter space where the model could deviate from \lcdm\ without a large likelihood penalty. Hence, early-universe new physics solutions to the Hubble tension are not ruled out, and are worth testing with \SO\ and future data.

\begin{table}[htbp]
\centering
\begin{tabular}{cccccc}
\hline
 & \lcdm\ & EDF & EDF & EDF, prior $\alpha=10^{-2}$ & EDF, prior $\alpha=10^{-3}$  \\
 & (NPIPE) & (NPIPE) & (2018 lite) & (2018 lite) & (2018 lite) \\
\hline
$d_1$ & - & $0.031_{-0.022}^{+0.012}$ & $<0.041$ & $<0.030$ & $<0.018$\\
$d_2$ & - & $<0.024$ & $<0.023$ &$<0.017$ & $<0.011$\\
$d_3$ & - & $<0.021$ & $<0.024$ & $<0.018$ & $<0.012$\\
$d_4$ & - & $<0.029$ & $<0.022$ & $<0.017$ & $<0.010$\\
$c_1^2$ & - & $0.46_{-0.16}^{+0.13}$ & $0.53_{-0.26}^{+0.24}$ & $0.54^{+0.29}_{-0.27}$ & unconstrained\\
$c_2^2$ & - & $0.30_{-0.06}^{+0.07}$ & $0.35_{-0.15}^{+0.06}$ & $0.38_{-0.38}^{+0.08}$ & unconstrained\\
$H_0$ & $67.3 \pm 0.5$ & $71.3_{-2.3}^{+1.6}$ & $69.8_{-1.8}^{+1.1}$ & $68.9_{-1.3}^{+0.9}$ & $68.2_{-1.1}^{+0.7}$\\
$\chi^2$ & $10963.7$ & $10960.0$ & - & - & -\\
$\Delta\textrm{AIC}$ & - & $-8.3$ & - & - & - \\
\hline
\end{tabular}
\caption{Constraints on the EDF model using \Planck\ data. We give the median value of the samples as well as the $68\%$ confidence interval. However, when providing upper limits, we give the $95\%$ confidence level interval. We also compare the best-fit sample $\chi^2$ with the \lcdm\ model and provide the $\Delta\textrm{AIC}$ value when comparing our model to \lcdm\ (in the case where we used the NPIPE likelihood). We additionally provide the constraints obtained when using Jeffreys-like priors on the density parameters in order to reduce the volume effects (see section~\ref{sec:jeffrey} for more details). In those cases, we made use of the \Planck\ 2018 lite likelihood, which allows us to more easily compare the posteriors with the mock \lcdm\ data. \label{tab:Planck}}
\end{table}

\subsection{Alternative priors}\label{sec:jeffrey}

\begin{figure}[htbp]
\centering
\includegraphics[width=\textwidth]{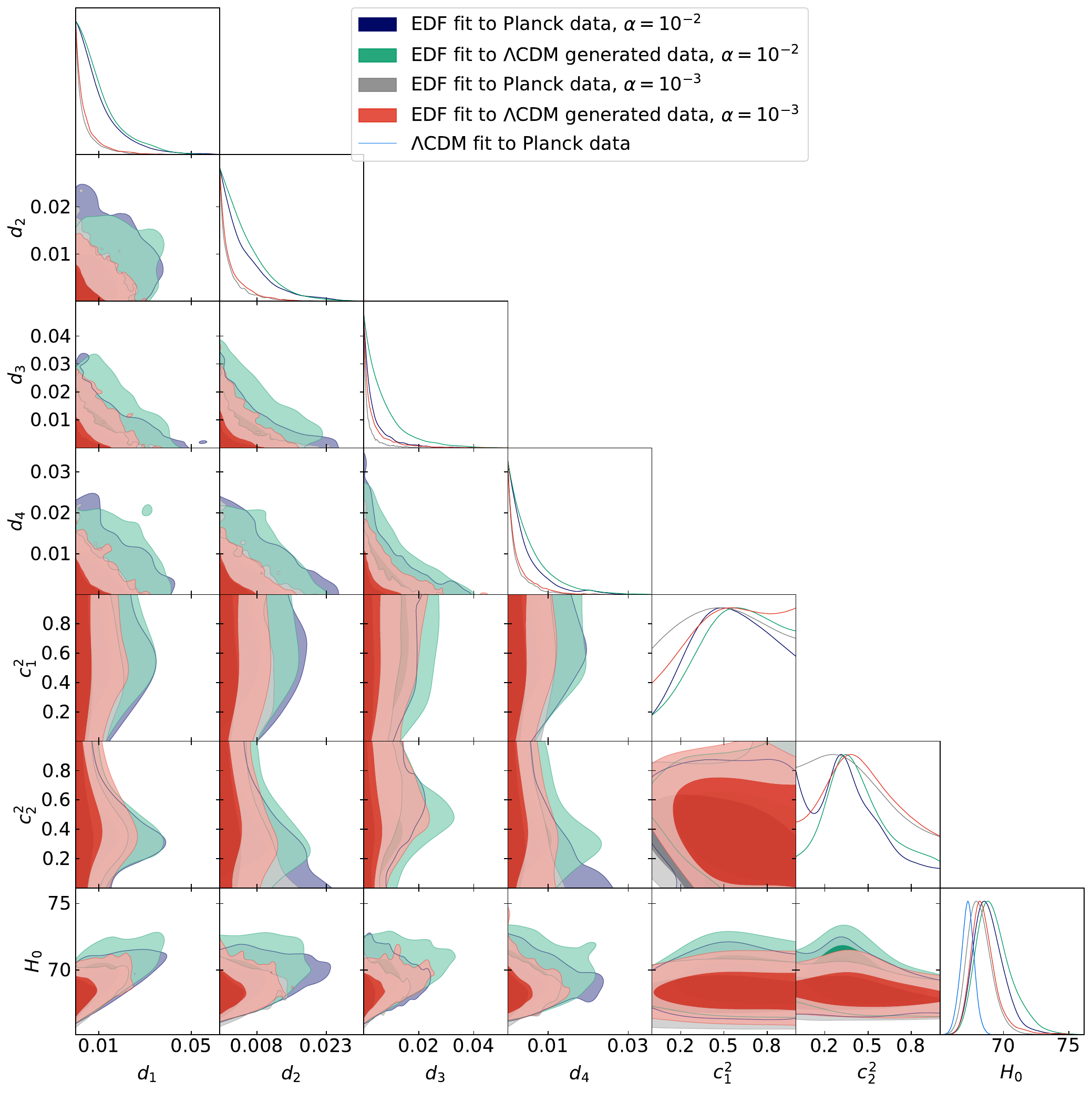}
\caption{Constraints on the six EDF parameters and $H_0$ when fitting our model to \Planck\ data with Jeffreys-like prior (dark blue and grey) or to \lcdm\ predicted power spectra with \Planck\ noise and Jeffreys-like prior (green and red). We also show the posterior of $H_0$ when fitting the \lcdm\ model to \Planck\ data (light blue). The noise always corresponds to the \Planck\ 2018 lite likelihood.\label{fig:triangle_plot_Jeffrey}}
\end{figure}

One way to mitigate these volume effects is by incorporating priors that better reflect our beliefs. We would like to use non-informative priors on the density amplitudes, which can be done by using Jeffreys-like prior. When a parameter $\theta$ represents a variance, Jeffreys prior takes the form $p=\frac{1}{\theta}$. Although our parameters do not represent variances, a similar simple prior can help to recover an $H_0$ posterior closer to the \lcdm\ posterior in the case where we fit the mock data generated with the \lcdm\ model. In practice, in order to avoid the singularity when $d_i=0$, we use a prior of the form,
\begin{align}
    p(d_i) = \frac{1}{d_i+\alpha},
\end{align}
where $\alpha$ is a regularization parameter set to be of the same order as $d_i$. We tested two different cases where $\alpha=10^{-2}$ or $\alpha=10^{-3}$. The constraints for these cases are shown in table~\ref{tab:Planck} and figure~\ref{fig:triangle_plot_Jeffrey}, where we have fitted our model to \Planck\ data (using the 2018 lite likelihood) with $\alpha=10^{-2}$ or $\alpha=10^{-3}$ (dark blue and grey contours), and compare the results with the mock data generated from the \lcdm\ fiducial model (green and red contours). As before, we also display the posterior of $H_0$ when fitting the \lcdm\ model to \Planck\ data (light blue). 

We observe in the two cases ($\alpha=10^{-2}$ and $\alpha=10^{-3}$) that the contours obtained when fitting the mock \lcdm\ data or the actual observations are very close to each other, emphasizing once more the consistency between \Planck\ data and the \lcdm\ model. Moreover, we see that adding the Jeffreys-like prior shifts the $H_0$ posterior closer to the \lcdm\ posterior (compared to figure~\ref{fig:triangle_plot_Planck}), especially when $\alpha=10^{-3}$. Similarly, as the results approach \lcdm, the posteriors for the sound speed parameters become flatter. The density parameters get more tightly constrained around zero, so larger deviations from \lcdm\ seem excluded.  A potential downside of using these priors is that models with likelihood values comparable to \lcdm\ are now disfavoured in marginalized constraints (as required to reduce the volume effect). This approach improves consistency for the $H_0$ and sound speed posteriors when the underlying model is \lcdm, but it also reduces the ability to detect new physics unless there is a strong impact on the CMB power spectra.

The conclusion of these tests is that \Planck\ data remains highly consistent with \lcdm, and we do not find strong preference for deviations, although the addition of a dark radiation component remains possible. Apart from $d_1$, the other modes are constrained to around $0.02-0.03$ at a $95\%$ confidence level, which still leaves room for the potential of exotic new physics. 

Although performing such an analysis is beyond the scope of this paper, it is worth noting that a profile likelihood analysis (for a few examples, see~\cite{Planck14,Chan18,Herold22,Herold24,Karwal24}) could also help mitigate volume effects. Profile likelihood analyses involve fixing a parameter of interest and maximizing the likelihood by varying all other parameters in the model. This approach allows one to construct confidence intervals for the parameter of interest that are independent of prior assumptions or parameter space volume considerations. As such, a profile likelihood analysis could complement MCMC methods, providing robust evidence for the detection of new physics.

\subsection{Understanding the \texorpdfstring{\lcdm\,--\,$d_1$}{LCDM-d1} degeneracy}\label{sec:phase_shift}

Finally, we try to understand better why, using \Planck\ data, $d_1$ is compatible with both zero (i.e. \lcdm\ when all other modes are also zero) and high values if the sound speed (especially around recombination) is close to $1/3$, which we refer to as the \lcdm\,--\,$d_1$ degeneracy.  

\begin{figure}[htbp]
\centering
\includegraphics[width=\textwidth]{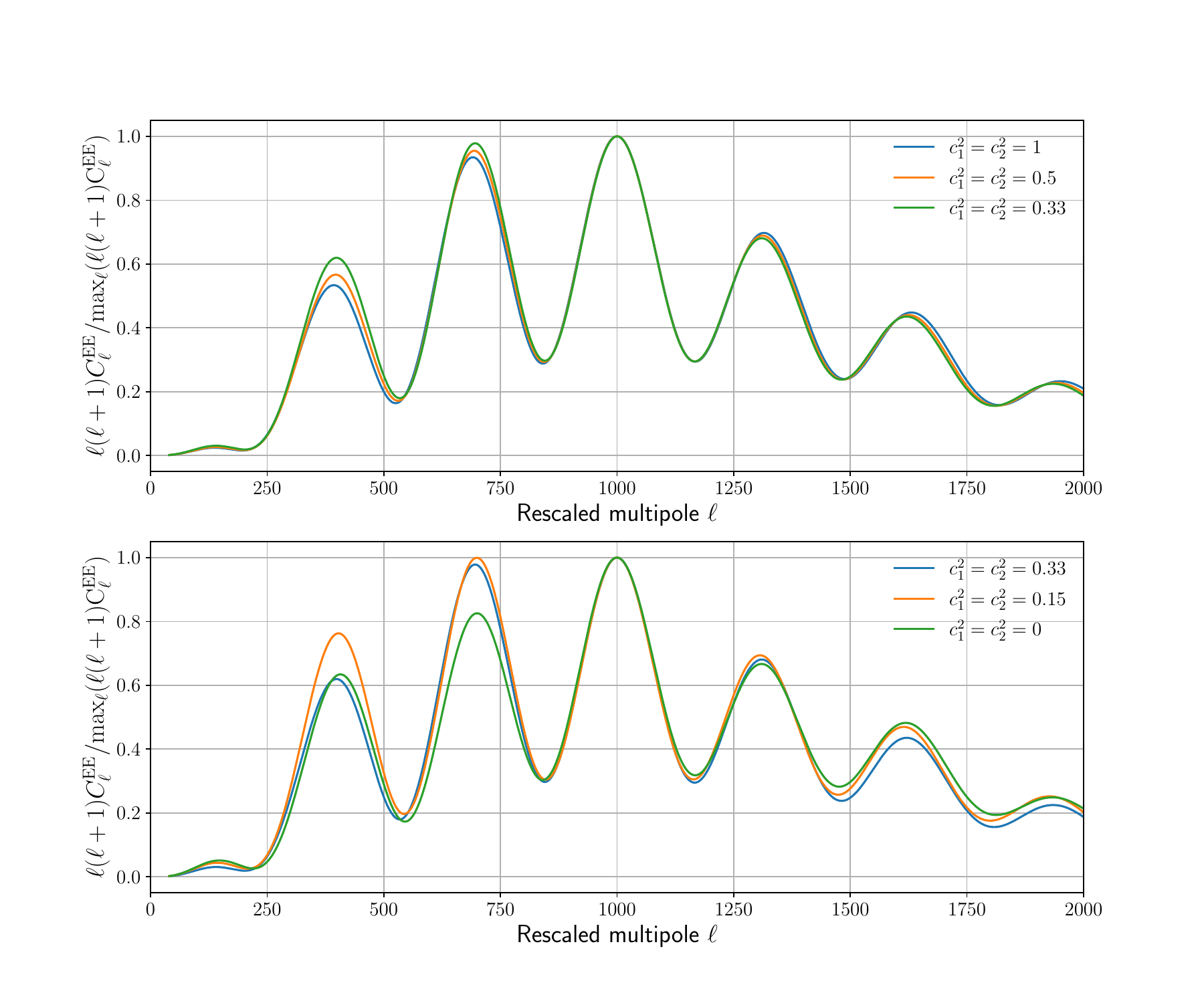}
\caption{Effect of varying the two sound speed parameters on the EE power spectra  with fixed background parameters and EDF density amplitudes $(d_1,d_2,d_3,d_4)=(0.15,0,0,0)$. We normalize the power spectra with respect to their maxima and rescale the multipoles so that their maxima are located at $\ell=1000$. Varying sound speed parameters in the range $[0.33,1]$ (upper panel) results in phase shifts, which cannot be fully compensated by rescaling multipoles. On the lower panel, the sound speed varies between $0$ and $0.33$ and the effect on the EE power spectra is a combination of phase shifts with larger relative changes in peak amplitudes. The effect of the sound speed on TT and TE power spectra is similar but can be seen better on the EE power spectrum. Note that the effect of the sound speed on the power spectra also depends on the density modes used; we here study the case where $d_1$ is nonzero as this is the mode for which \Planck\ data allows for the largest deviation from zero.}\label{fig:cl_ee_cs2}
\end{figure}

Setting $d_1=0.15$ and all other density modes to zero, we show the effect of varying the sound speed parameters on the EE power spectrum in figure~\ref{fig:cl_ee_cs2}, holding other parameters fixed. 
The amplitude and position of the peaks can be adjusted by varying $A_s$ and $\theta_*$ parameters, so we have normalized the power spectra with respect to their maximum, and their multipoles have been rescaled so that their maximum is located at $\ell=1000$. When the sound speed varies within the (approximate) range $[0.33,1]$, the acoustic peaks are systematically shifted towards larger scales as the sound speed increases, which comes from the fact that adding the EDF causes a phase shift at the perturbation level~\cite{Follin15,Baumann16}\footnote{For fixed background densities and observed peak positions, this means that higher sound speeds give fits with  lower $\theta_*$, where $\theta_*$ is defined using the background baryon-photon fluid sound speed.}. This phase shift is not strictly equivalent to a peak shift: as shown on the upper panel of figure~\ref{fig:cl_ee_cs2},  rescaling the multipoles  $\ell$ of the peaks  can only compensate for the phase shift at one specific scale. 

On the lower panel of figure~\ref{fig:cl_ee_cs2}, we show the cases where the sound speed varies within the range $[0,0.33]$. 
Here, the fluid can cluster as strongly as the other matter, 
and the effect on the power spectra mainly consists in a change of the relative heights of the acoustic peaks, combined with phase and other shifts. 

It therefore seems that a high $d_1$ value is allowed by \Planck\ data only in the case where the sound speed is close to $1/3$ because lower sound speeds have a very significant impact on the amplitude of the acoustic peaks which cannot be compensated by a change of $A_s$. Higher sound speeds are also excluded because of the phase shift they produce. As a result, it seems that adding a new component with a sound speed of $1/3$ makes it in phase with the other radiation, which is apparently still allowed with \Planck\ data.

Even though this model has entirely been developed to test deviations from \lcdm\ in the early Universe, using late-time probes could still be useful to constrain this model as they could help to break the degeneracies that may arise with \lcdm\ parameters. Additionally, late-time probes such as BAO or lensing measurements would also have nonzero yet limited sensitivity to our model since it can modify acoustic oscillations in the primordial plasma which leave an imprint in the galaxy distribution, and can also modify the matter power spectrum. In the case of the \lcdm\,--\,$d_1$ degeneracy, we use the minimum-variance baseline noise curve provided by the \SO\ collaboration and forecast that the reconstructed lensing power spectrum of~\SO\ would detect $d_1=0.05$ and $c_1^2=c_2^2=0.33$ at a significance of $6.4\sigma$ compared to \lcdm\ if all other parameters are known. This would not be that significant once all correlations and uncertainties of other parameters are taken into account, but it could still help to increase the constraining power of the data analysis. We use importance sampling to evaluate the impact of adding the lensing potential power spectrum on the $2\sigma$ upper-limit of the density amplitudes in the \lcdm\ fiducial case, finding that the constraints tighten by up to $6\%$, depending on the modes.

\section{Conclusion}

In this work, we developed a flexible parameterization for the properties of additional components in the early Universe, based on the GDM formalism, and tested its performance against both theoretical models and existing cosmological data from \Planck. Our primary goal in developing this model was to provide a framework for testing deviations from \lcdm\ using \SO\ observations. Additionally, the model offers the potential to suggest what type of new physics might be needed should a deviation from \lcdm\ be detected. By reproducing large classes of models, our parameterization could also help reduce computational resources by excluding several classes at once, streamlining the exploration of viable cosmological scenarios. We demonstrated that our model can effectively capture the effects of theoretical models like EDE and dark radiation at the level of the CMB power spectra. These models occupy distinct regions in the parameter space, making it easier to differentiate between them. However, distinguishing between specific EDE or dark radiation models may be more challenging due to the limited number of parameters we are using. Those two classes of models are mainly reproduced by the use of the $d_1$ and $d_3$ density amplitudes, which also suggests that $d_2$ and $d_4$, either independently or in combination with the other amplitudes, could represent as-yet-unknown models and provide solutions to the Hubble tension, especially with the flexibility to use different sound speeds.

Although our model does not outperform \lcdm\ when applied to the \Planck\ data, this result is expected since \lcdm\ fits \Planck\ well, and our model has many additional parameters. 
However, the data still allow for a large $d_1$ value and a sound speed around $1/3$, indicating something that behaves a bit like a dark radiation component. The other density modes are constrained to within a few percent, leaving room for new physics discoveries. Looking at the posterior of $H_0$, our model contains several parameter combinations that yield high values of $H_0$ that cannot be ruled out by \Planck\ and are worth exploring further with \SO. 

The application of our model to \Planck\ reveals significant volume effects. We found that applying priors can help mitigate these effects, but at the cost of reducing sensitivity to new physics. Our model was developed primarily for \SO, and has too many parameters to be able to cleanly separate fluctuations and new physics when using only \Planck\ data. The precision of \SO\ data will reduce the volume effects, and provide clearer constraints on deviations from \lcdm, offering the opportunity to either validate the standard model or uncover new physics in the early Universe. Indeed, although some volume effects may still be present with \SO, we showed that they could not be responsible for the inference of a high $H_0$ value, making any CMB-only solution to the Hubble tension using our model robust. 

Given the high number of additional parameters in our model, it is unlikely that a deviation from \lcdm\ will be detected with high significance through a single parameter. Instead, the detection of new physics should be based on comparing $\chi^2$ and AIC values, or more sophisticated methods that are left for future work. A significant improvement in $\chi^2$, independent of local $H_0$ measurements, would provide strong independent evidence for new physics.

\appendix
\section{Correlations between all parameters}\label{app:triangle_plot_lcdm_full_parameters}

\begin{figure}[htbp]
\centering
\includegraphics[width=\textwidth]{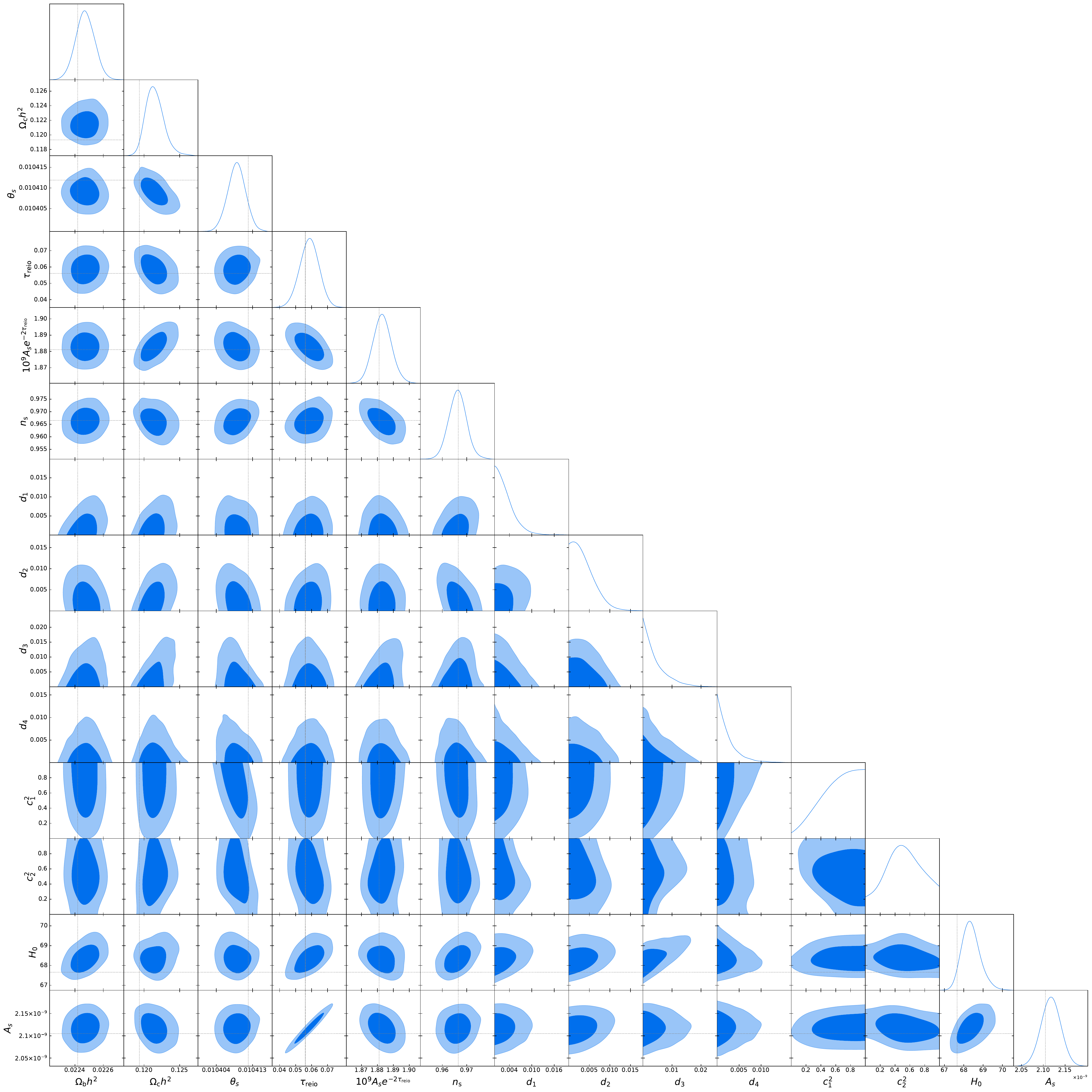}
\caption{Constraints on all the parameters varied in the model when fitting EDF to \lcdm\ fiducial power spectra with \SO\ noise. We also show the correlations with $H_0$ and $A_s$ although we did not use them as sampled parameters. Dashed grey lines depict the fiducial values used to generate the power spectra.\label{fig:triangle_plot_lcdm_full_parameters}}
\end{figure}

\begin{figure}[htbp]
\centering
\includegraphics[width=\textwidth]{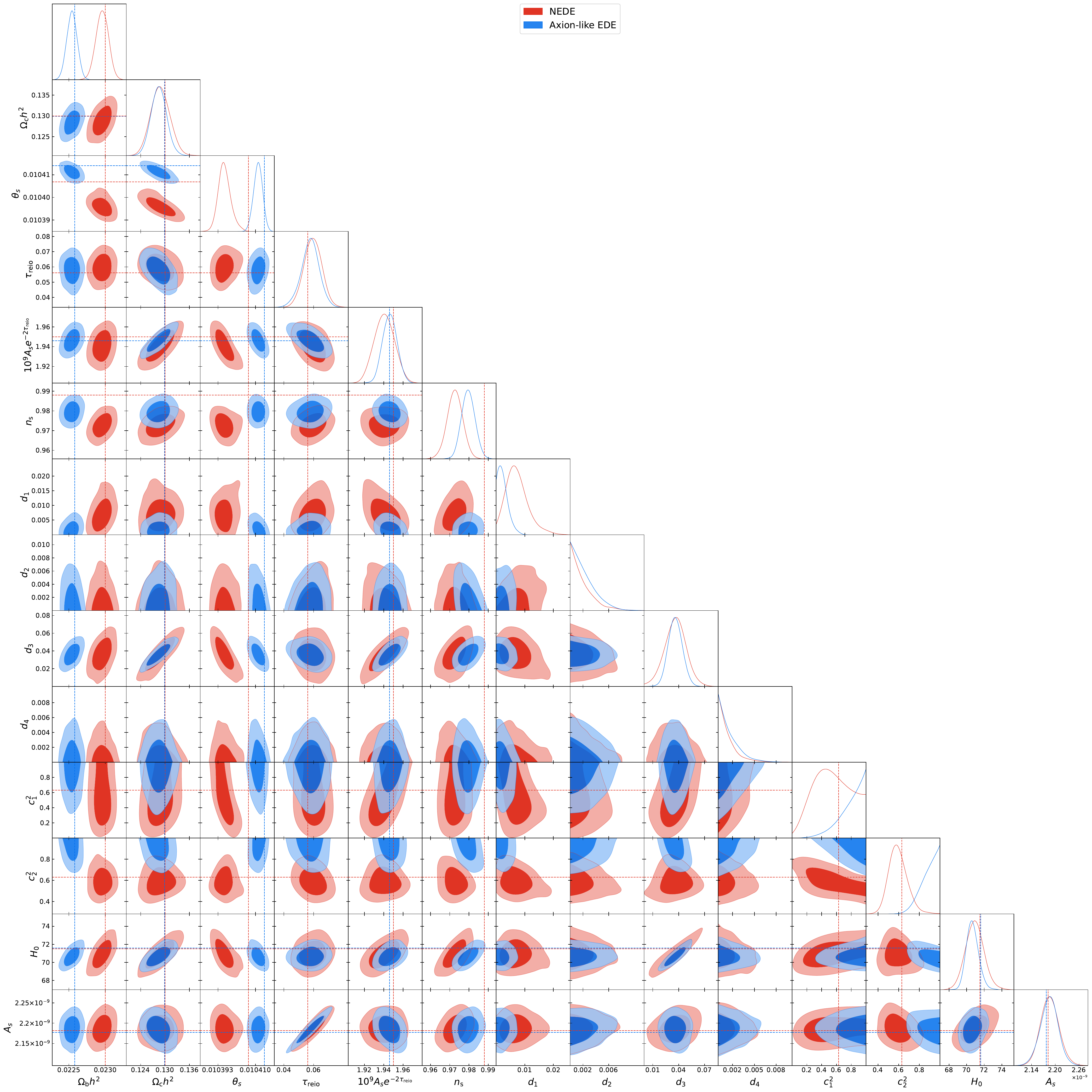}
\caption{Same as figure~\ref{fig:triangle_plot_lcdm_full_parameters} in the axion-like EDE (blue) or NEDE (red) cases.\label{fig:triangle_plot_EDE_full_parameters}}
\end{figure}

\begin{figure}[htbp]
\centering
\includegraphics[width=\textwidth]{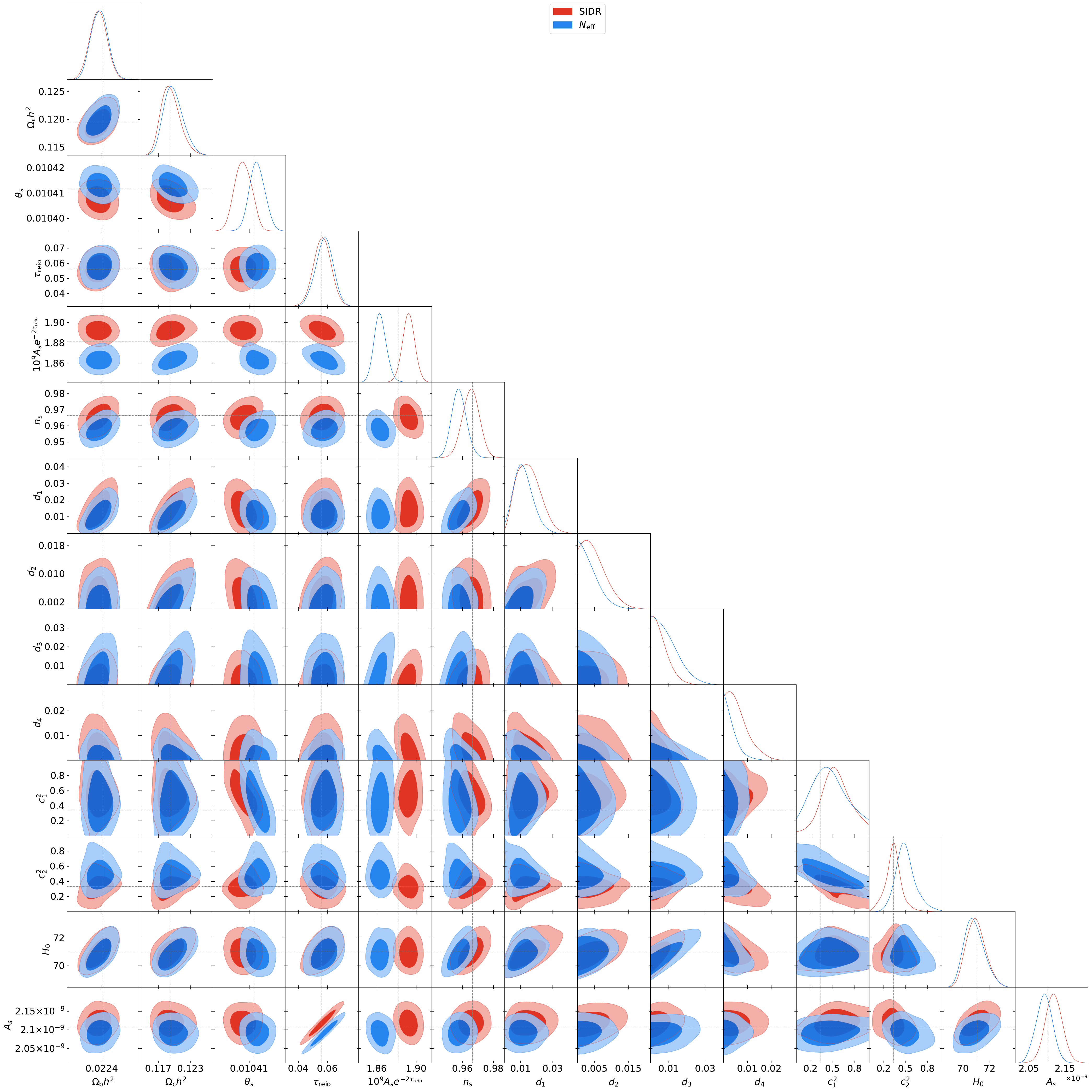}
\caption{Same as figure~\ref{fig:triangle_plot_lcdm_full_parameters} in the \Neff\ (blue) or SIDR (red) cases.\label{fig:triangle_plot_DR_full_parameters}}
\end{figure}

\acknowledgments

We are supported by UK STFC grant ST/X001040/1. We thank members of the SO collaboration for discussion.

% Bibliography

%% [A] Recommended: using JHEP.bst file
\bibliographystyle{JHEP}
\bibliography{biblio}
\end{document}